%% file: vdM_v13_prep.tex
\documentclass[ALICE,manyauthors]{cernphprep}
\usepackage[]{hyperref}
\usepackage{color}
\usepackage{xspace}
\begin{document}%
%%%%%%%%%%%%% ptdr definitions %%%%%%%%%%%%%%%%%%%%%
%
%%%%%%%%%%%%%%%  Title page %%%%%%%%%%%%%%%%%%%%%%%%
%
\begin{titlepage}
\PHnumber{2014-087}                 % required, obtained from PH
\PHdate{May 07, 2014}              % required
%\EXPnumber{ALICE-INT-2010-9999}     % optional
%\EXPdate{12 October 2010}           % optional
%
%
%%% Put your own title + short title here:
\title{Measurement of visible cross sections in proton-lead collisions at $\mathbf{\sqrt{s_{\rm{NN}}}}$~=~5.02~TeV in van der Meer scans with the ALICE detector}
\ShortTitle{Measurement of visible cross sections in proton-lead collisions at the LHC}   % appears on right page headers
%
%%% Do not change the next lines!
\Collaboration{ALICE Collaboration%
         \thanks{See Appendix~\ref{app:collab} for the list of collaboration
                      members}}
\ShortAuthor{ALICE Collaboration}      % appears on left page headers, do not change
\begin{abstract}
In 2013, the Large Hadron Collider provided proton-lead and lead-proton collisions at the center-of-mass energy per nucleon pair $\sqrt{s_{\rm{NN}}}$~=~5.02~TeV. Van der Meer scans were performed for both configurations of colliding beams, and the cross section was measured for two reference processes, based on particle detection by the T0 and V0 detectors, with pseudo-rapidity coverage \hbox{$4.6<\eta< 4.9$}, \hbox{$-3.3<\eta<-3.0$} and \hbox{$2.8<\eta< 5.1$, $-3.7<\eta<-1.7$}, respectively. Given the asymmetric detector acceptance, the cross section was measured separately for the two configurations. The measured visible cross sections are used to calculate the integrated luminosity of the proton-lead and lead-proton data samples, and to indirectly measure the cross section for a third, configuration-independent, reference process, based on neutron detection by the Zero Degree Calorimeters.
\end{abstract}
\end{titlepage}

\section{Introduction}

Luminosity determination in ALICE (A Large Ion Collider Experiment)~\cite{ALICE_jinst} at the LHC is based on visible cross sections measured in van der Meer (vdM) scans~\cite{vdM,balagura}. The visible cross section $\sigma_{\rm{vis}}$ seen by a given detector (or set of detectors) with a given trigger condition is a fraction of the total inelastic interaction cross section $\sigma_{\rm{inel}}$: $\sigma_{\rm{vis}}$~=~$\epsilon \sigma_{\rm{inel}}$, where $\epsilon$ is the fraction of inelastic events which satisfy the trigger condition.  In the following, an inelastic event satisfying a given trigger condition will be referred to as a reference process. Once  the reference-process cross section ($\sigma_{\rm{vis}}$) is measured, the collider luminosity can be determined as the reference-process rate divided by $\sigma_{\rm{vis}}$. This procedure does not require the knowledge of $\epsilon$.

In  vdM scans the two beams are moved across each other in the transverse directions $x$ and $y$. Measurement of the rate $R$ of the reference process as a function of the beam separation $\Delta x$, $\Delta y$ allows one to  determine the luminosity \textit{L} for head-on collisions of a pair of bunches with particle intensities $N_1$ and $N_2$ as
\begin{equation}
L = N_1 N_2 f_{\rm{rev}} / (h_x h_y)\rm{,}
\label{eq:Lumi}
\end{equation}
where $f_{\rm{rev}}$ is the accelerator revolution frequency and $h_x$ and $h_y$ are the effective beam widths in the  two transverse directions. The effective beam widths are measured as the area below the \hbox{$R(\Delta x,0)$} and \hbox{$R(0,\Delta y)$} curve (scan area), respectively, each divided by the head-on rate \hbox{$R(0,0)$}. Under the assumption that the beam profiles are Gaussian, the effective width is obtained as the Gaussian standard deviation parameter (from a fit) multiplied by $\sqrt{2\pi}$. However, the Gaussian assumption is not necessary for the validity of the method. As will be shown in section~\ref{sec:analysis}, other functional forms can be used, as well as numerical integration of the curve.
The cross section $\sigma_{\rm{vis}}$ for the chosen reference process is then
\begin{equation}
 \sigma_{\rm{vis}} = R(0,0)/L\rm{.}
 \label{eq:crSec}
 \end{equation} 

In 2013, the Large Hadron Collider provided proton-lead and lead-proton collisions at the center-of-mass energy per nucleon pair $\sqrt{s_{\rm{NN}}}$~=~5.02~TeV. Van der Meer scans were performed for both configurations of colliding beams, and the cross section was measured for two reference processes. In section~\ref{set-up}, the detectors used for the measurements are briefly described, along with the relevant machine parameters and the adopted scan procedure. In section~\ref{sec:analysis}, the analysis procedure is described. In section~\ref{sec:results}, the obtained results and uncertainties are presented and discussed. In  section~\ref{comparison}, the application of the vdM scan results to the measurement of the integrated luminosity is briefly discussed. In section~\ref{zdc}, the vdM scan results are used to indirectly determine the cross section for a third reference process, based on neutron detection by the ALICE Zero Degree Calorimeters.

\section{Experimental setup}\label{set-up}

At the ALICE experiment, two vdM-scan sessions were carried out  during the 2013 proton-lead data-taking campaign at the LHC. The proton beam was travelling clockwise in the first session and counter-clockwise in the second session. In the following, these configurations will be referred to as p--Pb and Pb--p, respectively.  

In each session, the cross section was measured for two reference processes: one is based on the V0 detector, the other on the T0 detector. A detailed description of these detectors is given in~\cite{ALICE_jinst}, and their performance is discussed in~\cite{aliPerf}, \cite{tzeroPerf} and~\cite{vzeroPerf}. The V0 detector consists of two hodoscopes, with 32 scintillator tiles each, located on opposite sides of the ALICE Interaction Point (IP2), at a distance of 340~cm (V0-A) and 90~cm (V0-C) along the beam axis, covering the pseudo-rapidity ($\eta$) ranges $2.8<\eta< 5.1$ and $-3.7<\eta<-1.7$, respectively. In the p--Pb configuration the proton beam is travelling in the direction from V0-A to V0-C. 
The T0 detector consists of two arrays of 12 Cherenkov counters each, located on opposite sides of IP2, at a distance of 370~cm (T0-A) and 70~cm (T0-C) along the beam axis, covering  the pseudo-rapidity ranges \hbox{$4.6<\eta< 4.9$} and \hbox{$-3.3<\eta<-3.0$}, respectively. In the p--Pb configuration the proton beam is travelling in the direction from T0-A to T0-C.

The V0-based trigger condition, chosen as the reference process,  requires at least one hit in each detector hodoscope, i.e. on both sides of IP2. As discussed in~\cite{aliMult}, the efficiency of such a selection is larger than 99\% for non single-diffractive p--Pb collisions. A similar trigger condition defines the T0-based reference process, with the additional condition that the longitudinal coordinate of the interaction vertex, evaluated by the trigger electronics via the difference of arrival times  in the two arrays (measured with a resolution of 20~ps), lies in the range $|z| <$~30~cm (where $z$~=~0 is the nominal IP2 position). This online cut aims to reject the background  from beam-gas and beam-satellite interactions. The cut value of 30 cm is much larger than the r.m.s. longitudinal size of the interaction region ($\simeq$~6~cm), making signal loss induced by the cut negligible ($<$10$^{-5}$). 
Since the two LHC beams have the same magnetic rigidity and different projectile mass, the energy per nucleon of lead ions (1.58 TeV) differs from that of protons (4 TeV). Hence, the p--Pb (\hbox{Pb--p}) collision center-of-mass frame is shifted by 0.47 (-0.47) units of rapidity with respect to the ALICE frame. Due to this shift and the asymmetric setup of both detectors, there is no reason to expect identical cross sections for the p--Pb and Pb--p configurations. Therefore, the results obtained in the two scan sessions are not combined.

\begin{figure}[tbp]
\begin{center}
\includegraphics[width=0.48\textwidth]{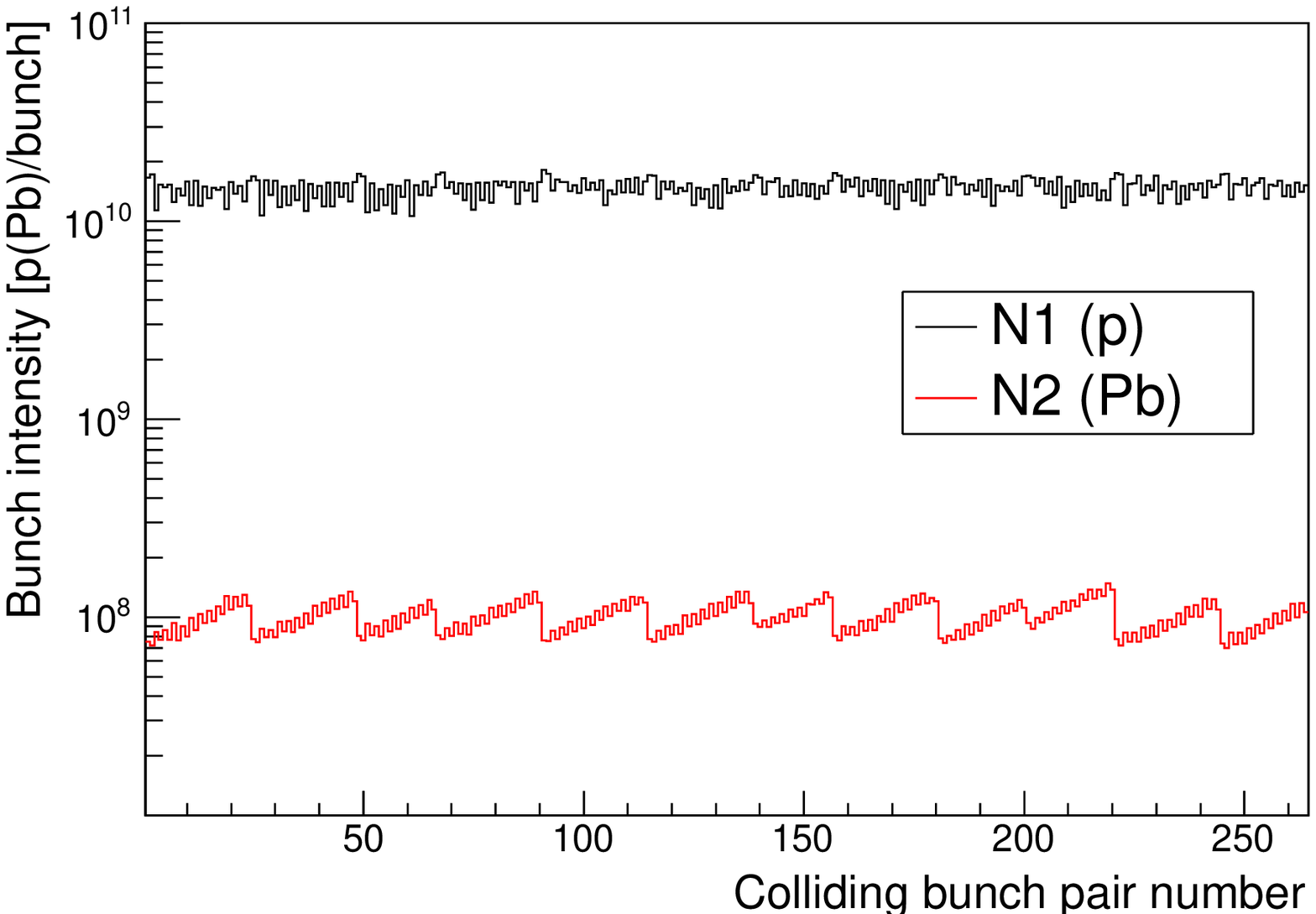} 
\includegraphics[width=0.48\textwidth]{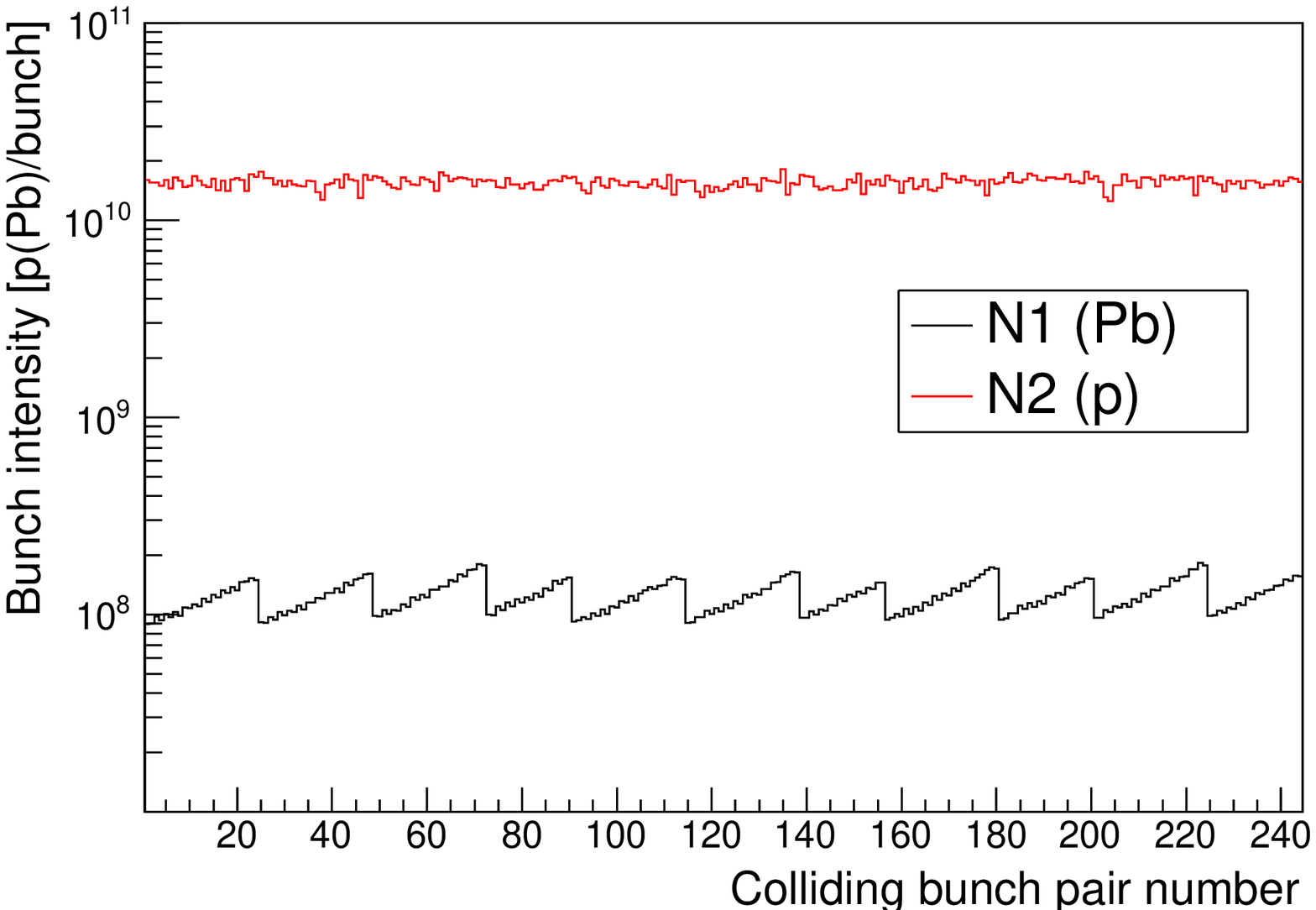} 
\end{center}
   \caption{(Colour online) Bunch intensities $N_1$ and $N_2$ for all colliding bunches, for an arbitrary timestamp during the p--Pb (left) and Pb--p (right) scan sessions.}
\label{fig:bunchInt}
\end{figure}

In the p--Pb (Pb--p) scan session the proton beam consisted of 272 bunches, while the Pb beam consisted of 338 (314) bunches. In the p--Pb (Pb--p) scan session 264 (244) bunch pairs per LHC orbit were colliding at IP2. For both beams and sessions, the minimum spacing between two consecutive bunches was 200~ns. The reference-process rates were recorded (and the cross section measured) separately for each colliding bunch pair. For each session, two independent measurements per bunch pair were performed by repeating the (horizontal and vertical) scan pair twice: from negative to positive separation and then in the opposite direction. The maximum beam separation during the scan was about 0.15~mm, corresponding to about six times the RMS of the transverse beam profile.  In both sessions, the $\beta^*$ value\footnote{The $\beta(z)$ function describes the single-particle motion and determines the variation of the beam envelope as a function of the coordinate along the beam orbit ($z$). The notation $\beta^*$ denotes the value of the $\beta$ function at the interaction point.} in IP2 was 0.8~m. The current in the ALICE solenoid (dipole) was 30~kA (6~kA),  corresponding to a field strength of 0.5~T (0.7~T). In both sessions, the proton and lead bunch intensities were on the order of 10$^{10}$ p/bunch and 10$^{8}$ Pb/bunch, as shown in figure~\ref{fig:bunchInt}. While the proton bunch intensity is reasonably constant across bunches, large variations are seen for the lead bunches. The structure of such variations as a function of the bunch position can be explained by  different sensitivities to losses in the injection chain~\cite{schaumann}.    

The bunch-intensity measurement is provided for both scan sessions by the LHC instrumentation~\cite{beam_current}: a DC current transformer (DCCT), measuring the total beam intensity, and a fast beam current transformer (fBCT), measuring the relative bunch populations. The measured beam intensity is corrected by the fraction of ghost  and satellite charge\footnote{The radio-frequency (RF) configuration of the LHC is such that the accelerator orbit is divided in 3564 slots of 25~ns each. Each slot is further divided in ten buckets of 2.5~ns each. In nominally filled slots, the particle bunch is captured in the central bucket of the slot. Following the convention established in~\cite{bcnwg_long}, the charge circulating outside of the nominally filled slots is referred to as ghost charge;  the charge circulating within a nominally filled slot but not captured in  the central bucket is referred to as satellite charge.}. The measurement of ghost charge is provided by the LHCb collaboration, via the rate of beam-gas collisions occurring in nominally empty bunch slots, as described in~\cite{c_barschel_phd}. The obtained ghost-charge correction factor to the bunch-intensity product $N_1N_2$ is 0.991$\pm$0.001 (0.986$\pm$0.002) for the p--Pb (Pb--p) session. The bunch intensity is further corrected by the fraction of satellite charge measured by the LHC Longitudinal Density Monitor (LDM), which measures synchrotron radiation photons emitted by the beams~\cite{ldm_note}. The obtained satellite-charge correction factor to the bunch-intensity product $N_1N_2$ is 0.998$\pm$0.004 (0.996$\pm$0.001) for the p--Pb (Pb--p) session. This correction is implemented by multiplying the $N_1N_2$ product by both the ghost- and satellite-charge factors.

\begin{figure}[tbp]
\begin{center}
\includegraphics[width=0.48\textwidth]{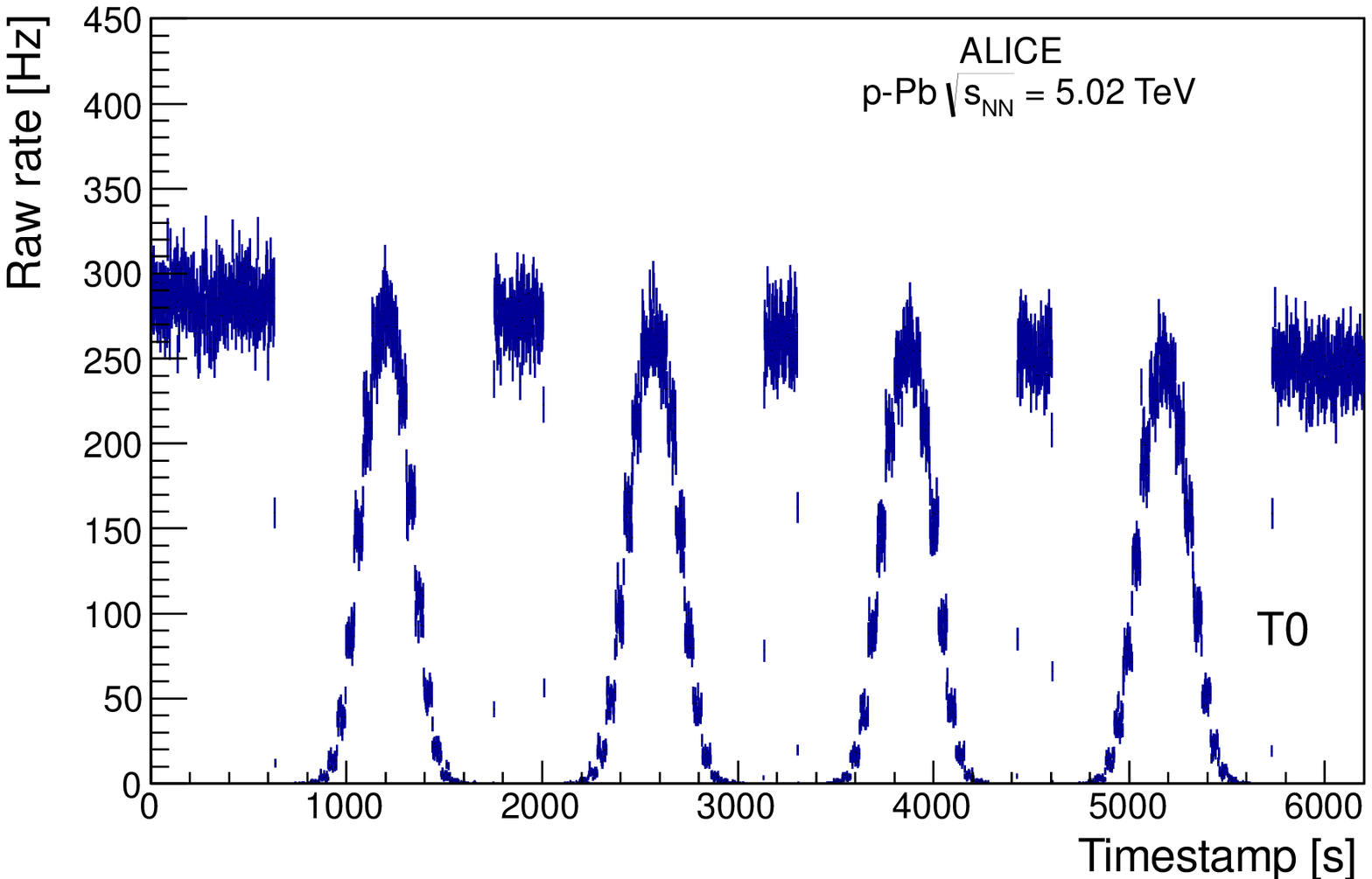} 
\includegraphics[width=0.48\textwidth]{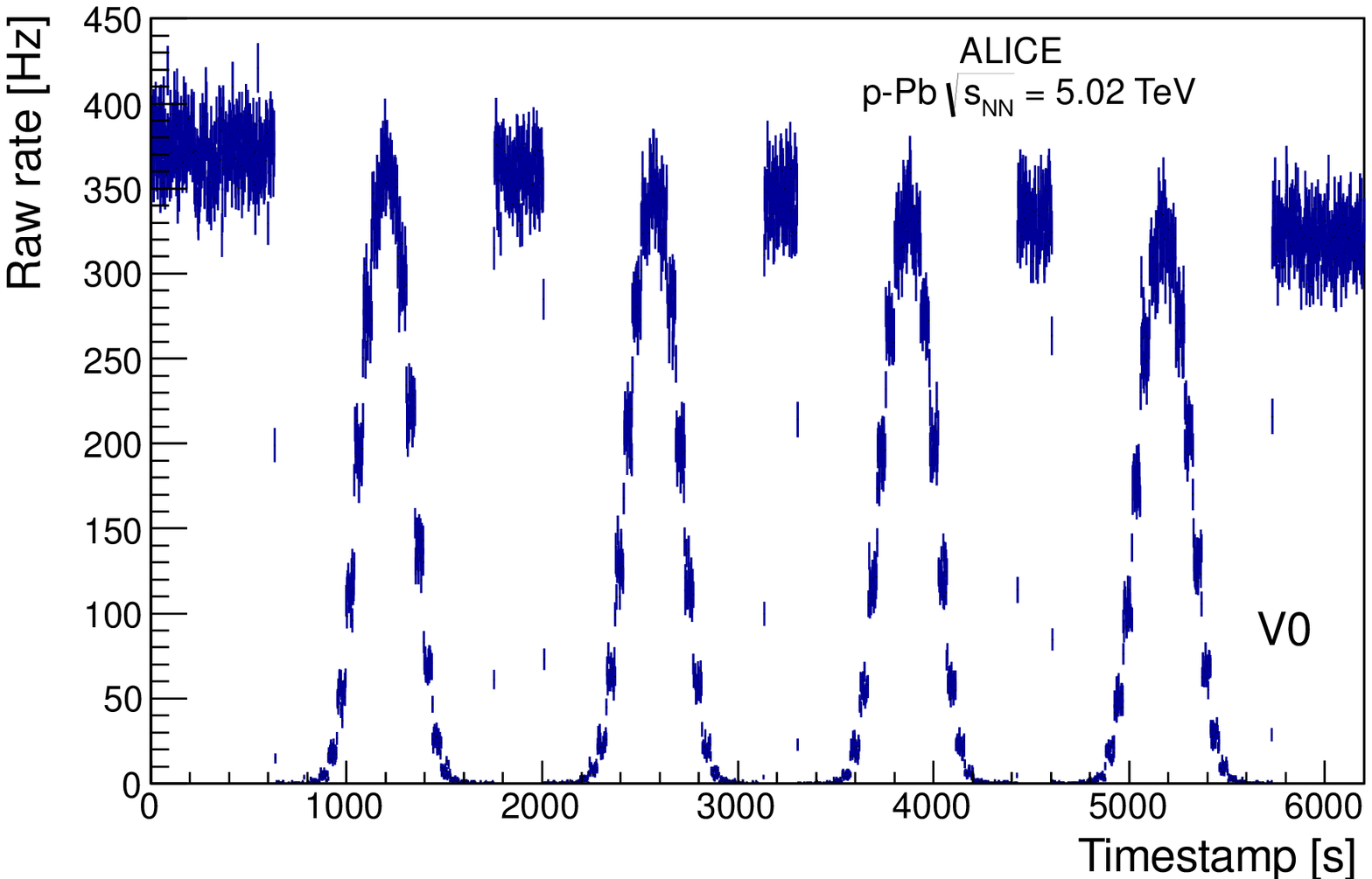} \\
\end{center}
   \caption{(Colour online)  Raw rate of the T0 (left) and V0 (right) process for a typical colliding bunch pair, as a function of time, during the p--Pb scan session. In each plot, the first (second) bell-shaped structure corresponds to the beam separation in the horizontal (vertical) direction being varied from negative to positive values. The third (fourth) bell-shaped structure corresponds to the beam separation in the horizontal (vertical) direction being varied from positive to negative values.}
\label{fig:RawRate}
\end{figure}

\section{Data analysis} \label{sec:analysis}

An example of the measured raw rate for one typical pair of colliding bunches during the p--Pb scan is shown in figure~\ref{fig:RawRate} for both the T0- and the V0-based processes.

\begin{figure}[tbp]
\begin{center}
\includegraphics[width=0.48\textwidth]{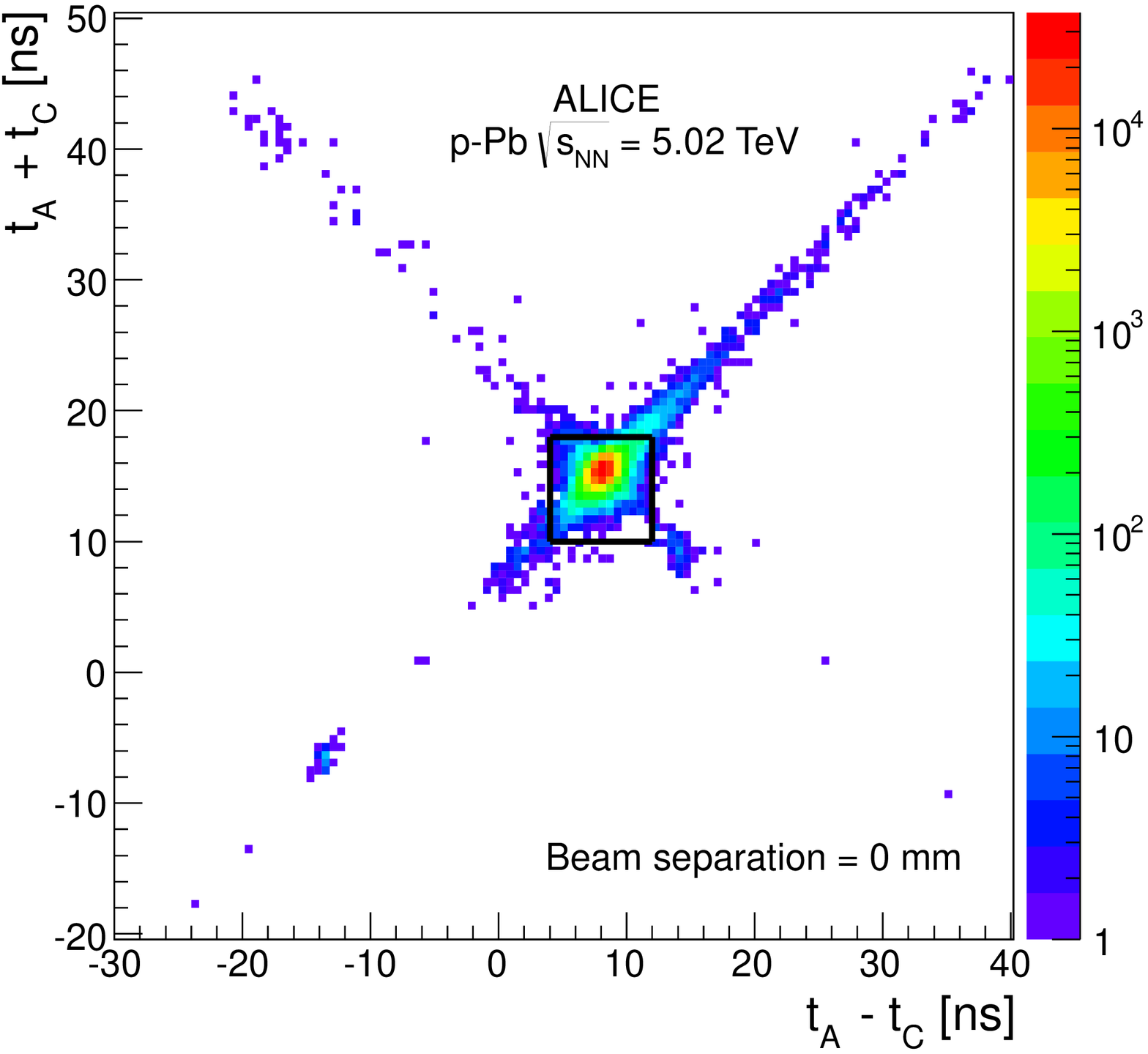} 
\includegraphics[width=0.48\textwidth]{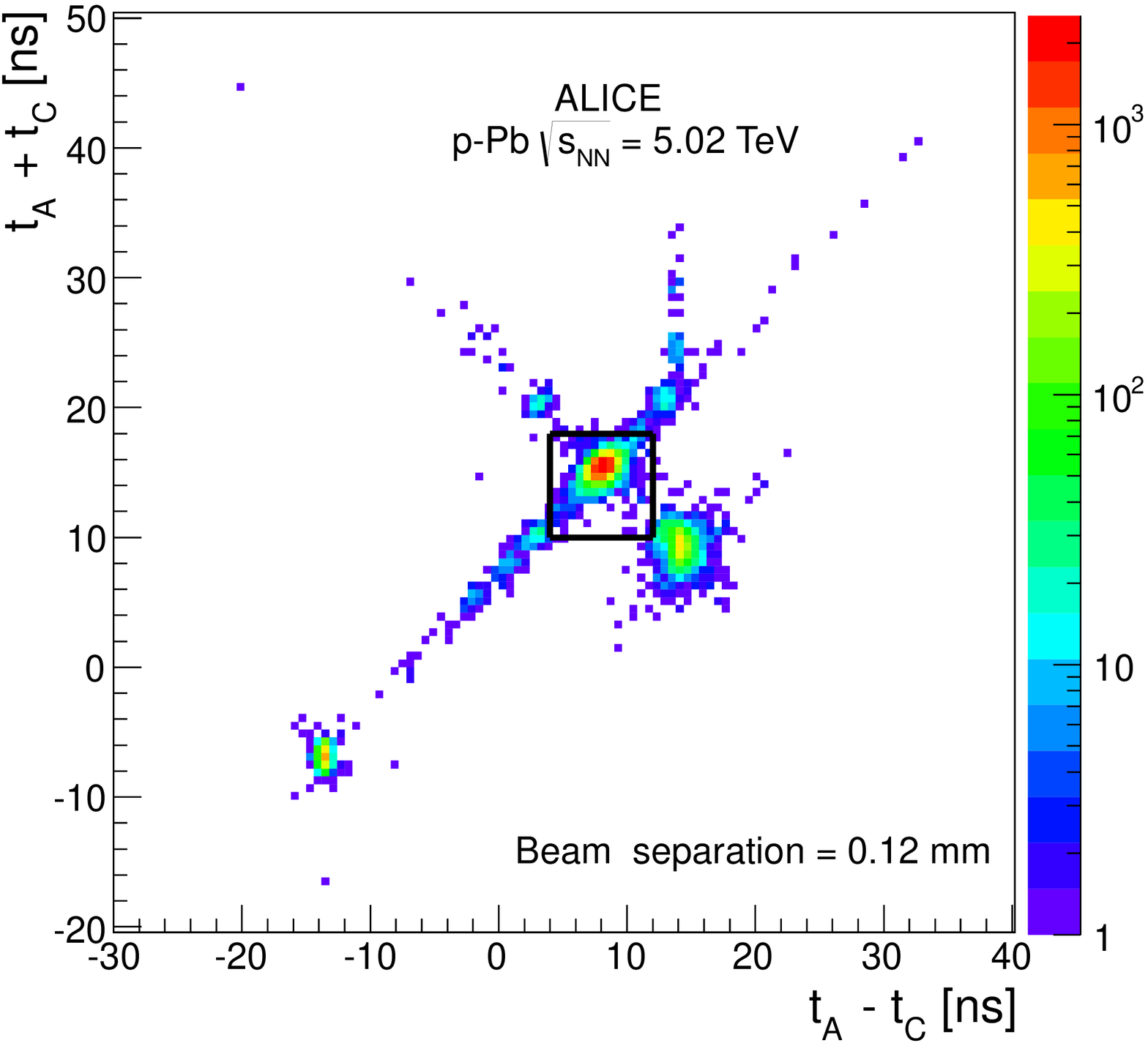} \\
\end{center}
   \caption{(Colour online) Correlation between the sum and difference of arrival times (relative to the bunch crossing) on the two V0 arrays. The left plot was obtained at zero beam separation; the right plot was obtained at a beam separation of 0.12~mm, roughly corresponding to five times the RMS of the beam profile. Events lying inside the area within the continuous lines are flagged as beam-beam interactions.}
\label{fig:bkg}
\end{figure}

Three  corrections are applied to the measured raw rate for each of the two reference processes.

First,  the contamination by beam--satellite and beam--gas interactions in the V0 rate is removed using the detector timing capabilities. The background is identified via the sum and difference of arrival times in the two V0 arrays from offline analysis of the data collected during the scan. The arrival times are obtained by averaging over the signal times of all hits of each array. The background contamination is measured as the fraction of events in which the sum and difference of times  lie outside of a window of $\pm$4~ns around the values expected for beam-beam collisions (figure~\ref{fig:bkg}).  The measurement is performed for each separation value and the corresponding raw rate is corrected by the obtained fraction. The background contamination in the V0-triggered sample is about 0.5-1\% at zero separation and about 20-40\% at a separation corresponding to five times the beam RMS. This procedure has negligible effect ($<$~0.1\%) when applied  to the T0 rates, due to the vertex cut in the T0 trigger logic described in section~\ref{set-up}. 
In order to study a possible contamination of the trigger rate by the intrinsic noise counts of the detectors, the rate of both trigger signals in absence of beam was measured and found to be zero. The rate in empty bunch slots with beam circulating was also measured and found to be zero for T0. For V0, a non-zero rate is measured up to the fourth empty bunch slot after a filled slot. Since the minimum spacing between filled slots is eight slots, such an after-pulsing effect does not affect the measurement of the rate in colliding slots.

Second, the probability of multiple interactions in the same bunch crossing (pileup) is taken into account according to Poisson statistics. 
The trigger rate $R$ is smaller than the rate  of visible interactions  by a factor
$[1-\exp{(-\mu_{\rm{vis}}})]/\mu_{\rm{vis}}$, where $\mu_{\rm{vis}}$~=~-$\ln(1-R/f_{\rm rev})$ is the average number of visible interactions occurring in one bunch crossing.
The pileup-corrected rate for bunch crossing $i$,  $R_{{\rm PU},i}$, is thus given by 
\begin{equation}
R_{{\rm PU},i} = -f_{\rm rev}\;\ln(1-R_{{\rm BB},i}/f_{\rm rev})
\end{equation}
where $R_{{\rm BB},i}$ is the background-corrected rate. In both scan sessions, the maximum value of $\mu_{\rm{vis}}$ during the scan for the V0 (T0) reference process is about 0.05 (0.03), leading to a maximum correction of about 2.5\% (1.5\%). 

\begin{figure}[tbp]
\begin{center}
\includegraphics[width=0.48\textwidth]{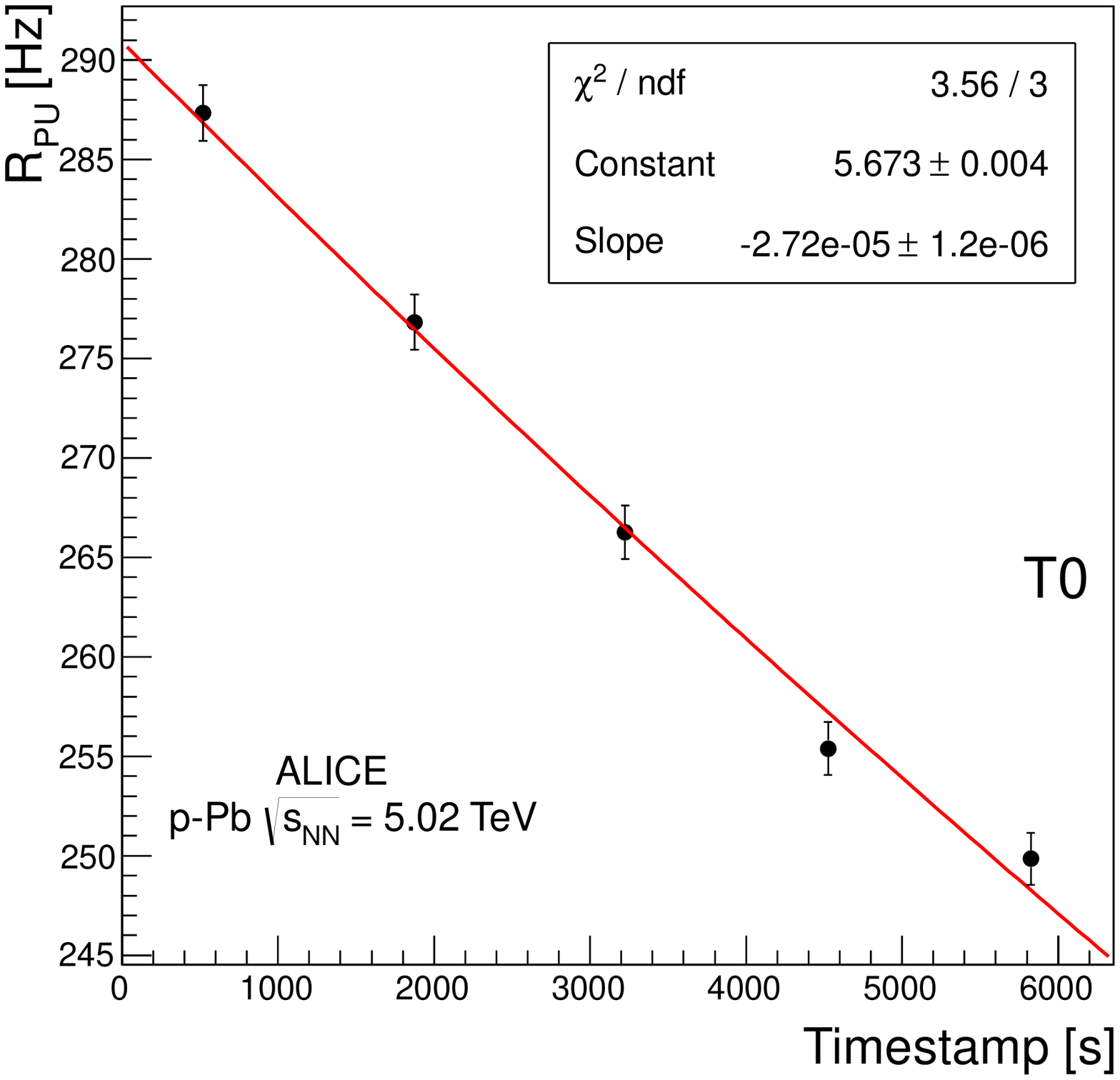} 
\includegraphics[width=0.48\textwidth]{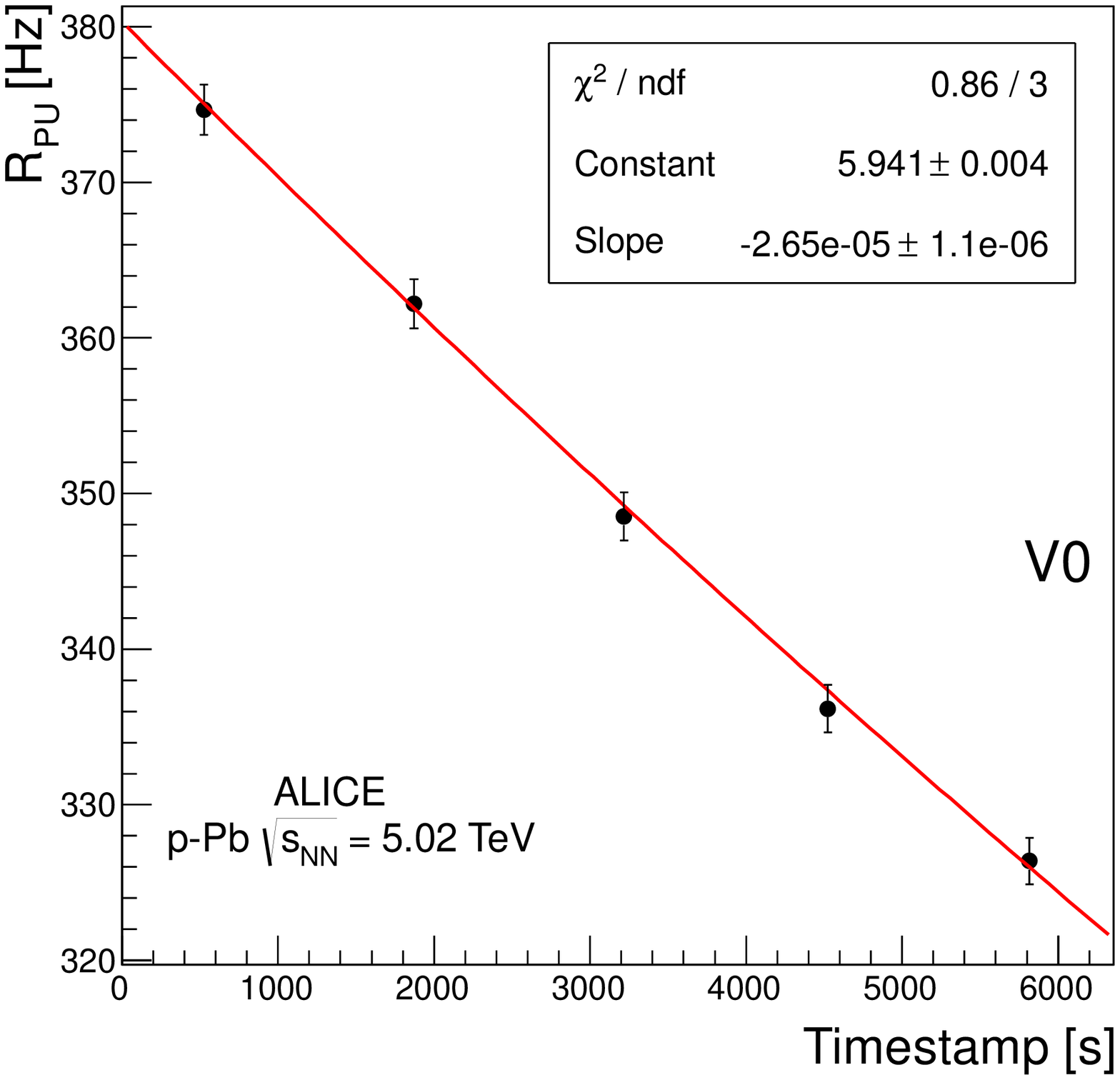} 
\end{center}
   \caption{(Colour online)  Background- and pileup-corrected head-on rates of the T0 (left) and V0 (right) reference process as a function of time for one interacting bunch crossing in the p--Pb scan session. The solid red curve is an exponential fit to the data points.}
\label{fig:ExpFit}
\end{figure}

The third correction takes into account that the luminosity decreases with time (as can be seen in figure~\ref{fig:RawRate}) due to the beam-intensity  decay and to the growth of emittances. In order to correct for this effect,  the evolution of the head-on luminosity in time is parameterised via a fit to the rates at zero separation measured before, after and in-between scans. The decay rate is satisfactorily described by an exponential function. Figure \ref{fig:ExpFit} shows an example of such a fit. The obtained fit parameters are used to normalize all rates of a given scan pair to an arbitrary reference time, chosen to lie between the horizontal and vertical scans. 

\begin{figure}[tbp]
\begin{center}
\includegraphics[width=0.49\textwidth]{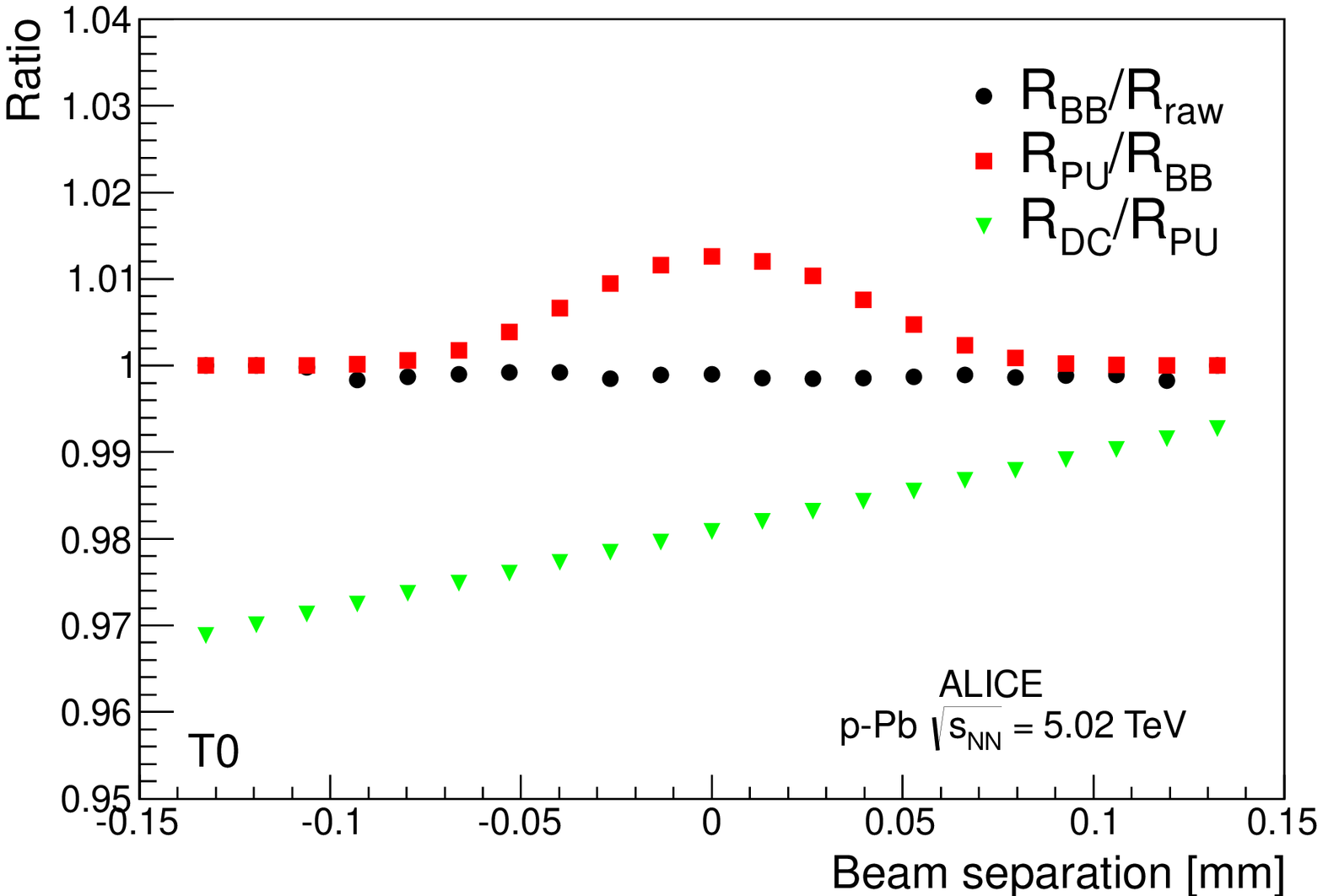} 
\includegraphics[width=0.49\textwidth]{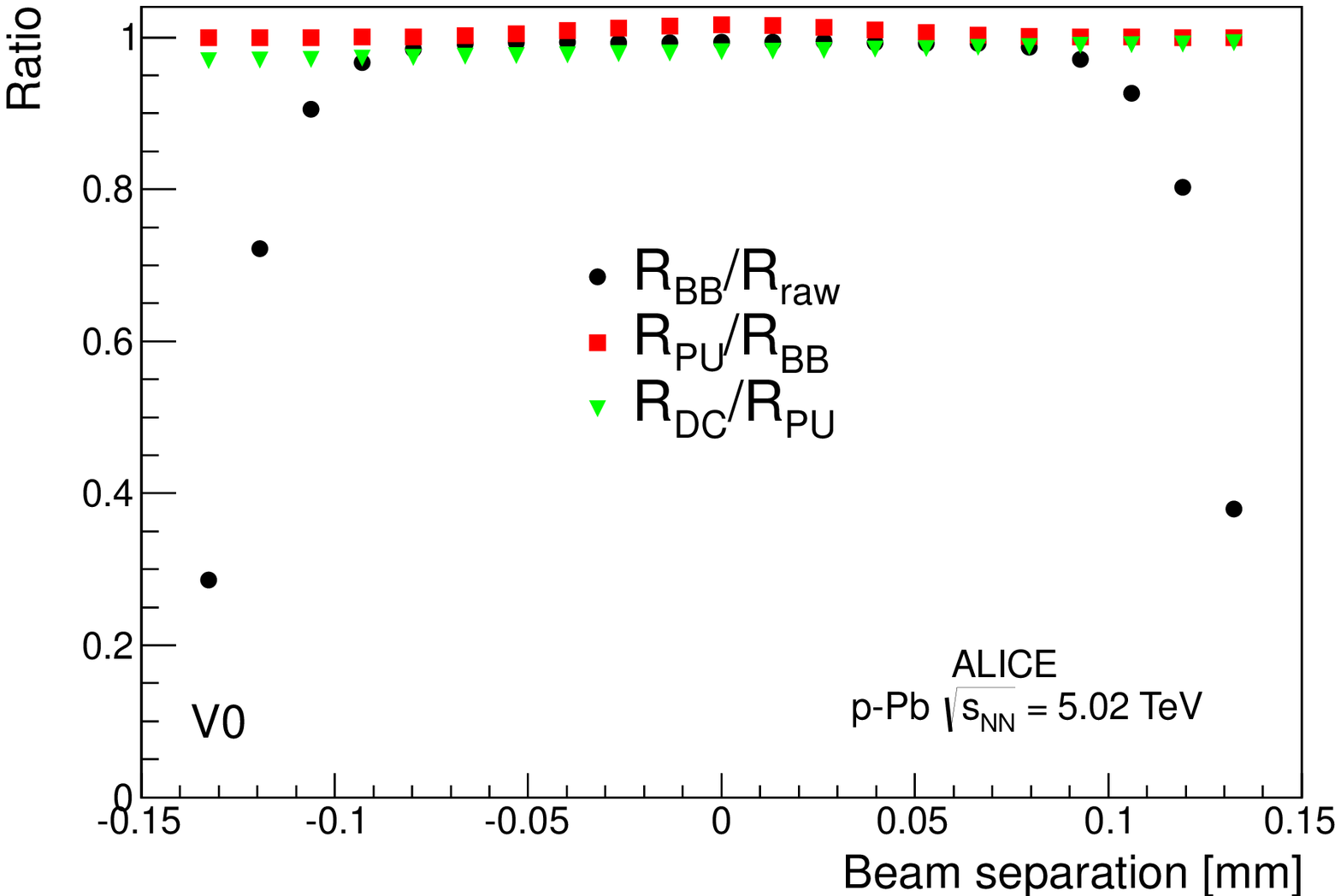} \\
\end{center}
   \caption{(Colour online)  Background (R$_{\rm{BB}}$/R$_{\rm{raw}}$), pileup (R$_{\rm{PU}}$/R$_{\rm{BB}}$) and luminosity decay (R$_{\rm{DC}}$/R$_{\rm{PU}}$) correction factors to the T0 (left) and V0 (right) rates as a function of the beam separation  for one typical pair of colliding bunches during the first p--Pb vertical scan. Due to the different size of the background correction factor for T0 and V0, the two figures have different vertical scales.}
\label{fig:RateCorr}
\end{figure}

An example of the obtained correction factors as a function of the beam separation  is shown in figure~\ref{fig:RateCorr}.

\begin{figure}[tbp]
\begin{center}
\includegraphics[width=0.48\textwidth]{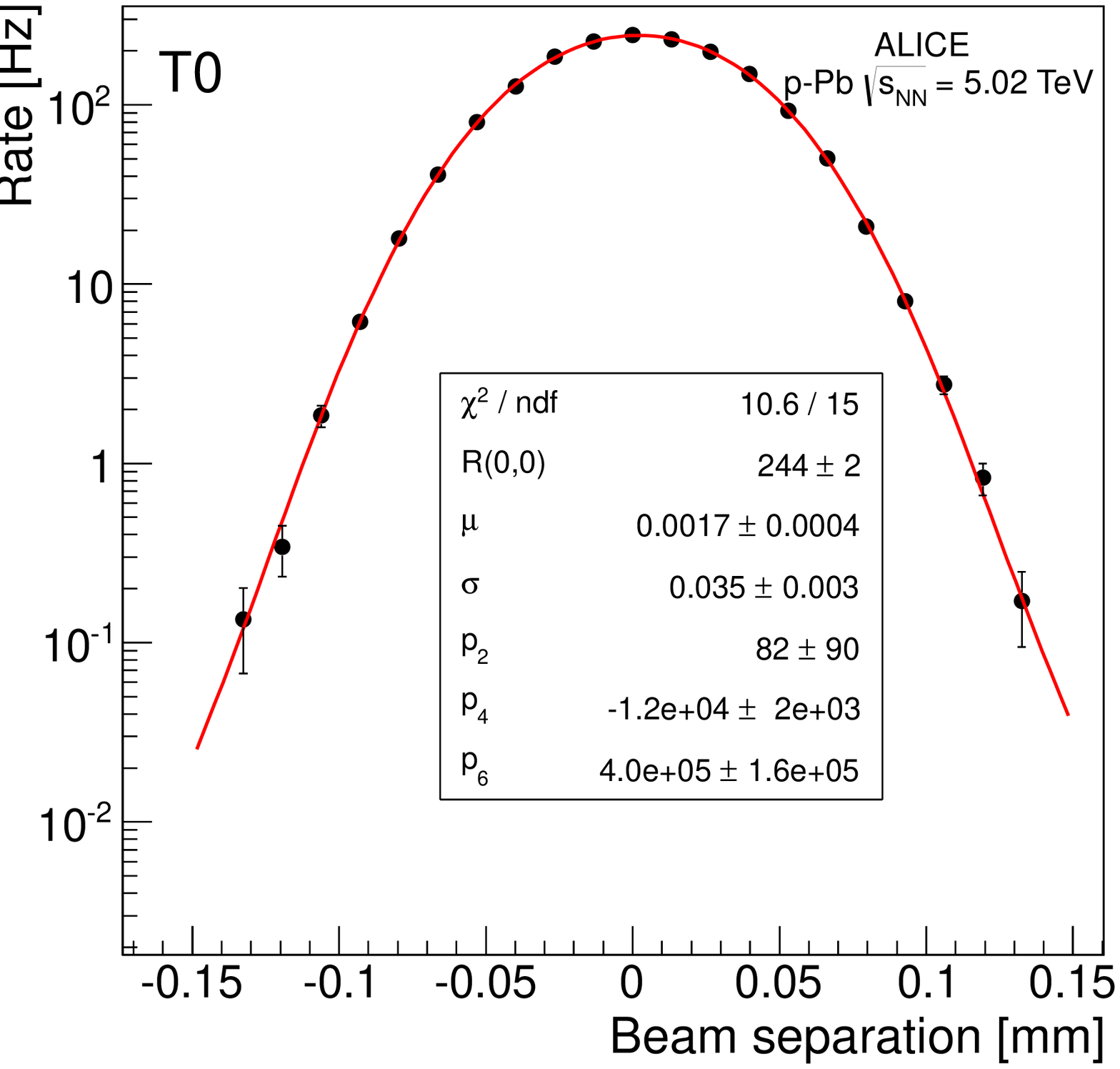} 
\includegraphics[width=0.48\textwidth]{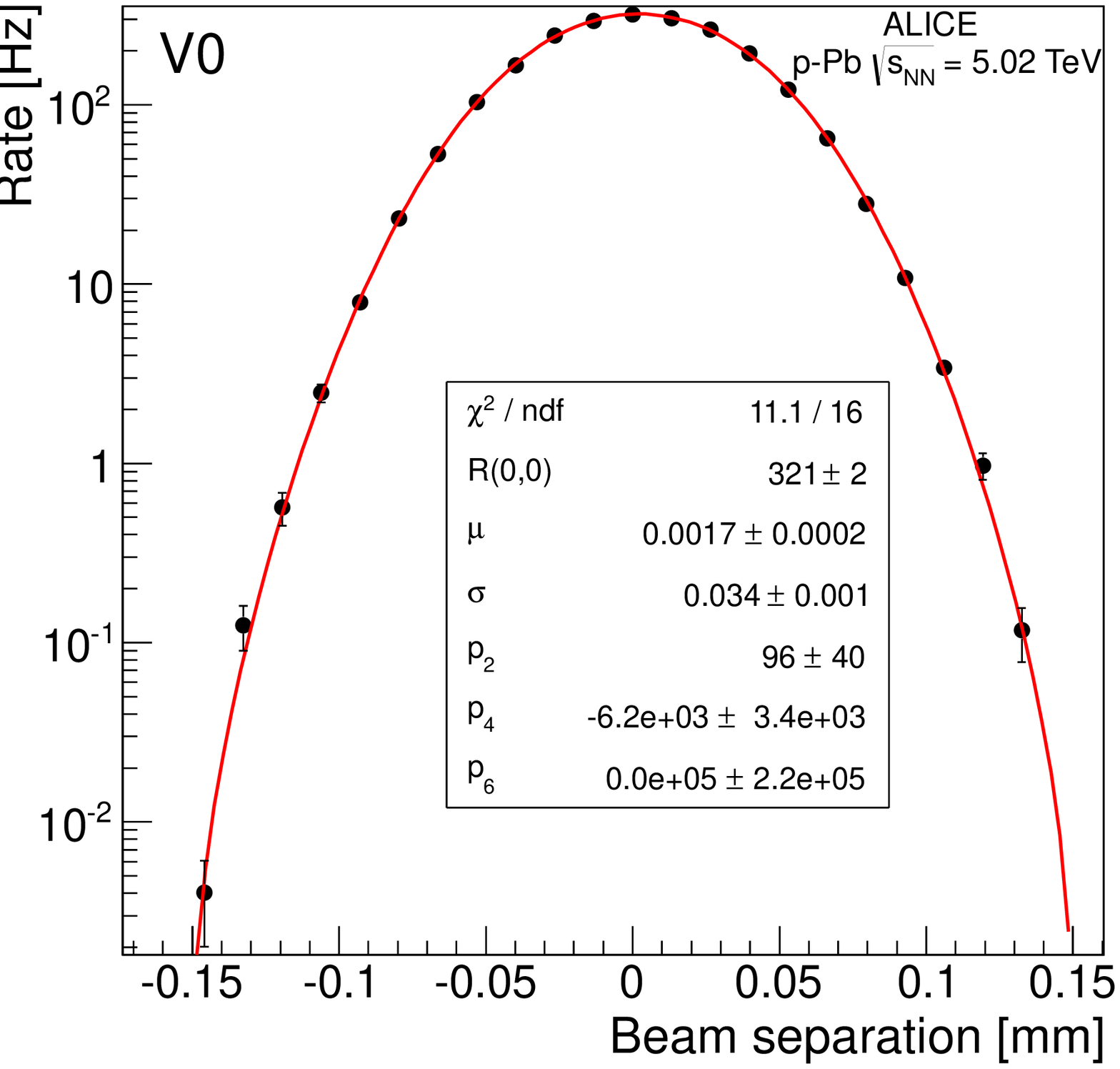} \\
\end{center}
   \caption{(Colour online)  Rates of the T0 (left) and V0 (right) reference process as a function of beam separation for one typical pair of colliding bunches in the first vertical p--Pb scan. The solid red curve is a fit according to eq.~\protect \ref{eq:fit}.}
\label{fig:ShapeFit}
\end{figure}

\begin{figure}[tbp]
\begin{center}
\includegraphics[width=0.49\textwidth, height=0.52\textwidth]{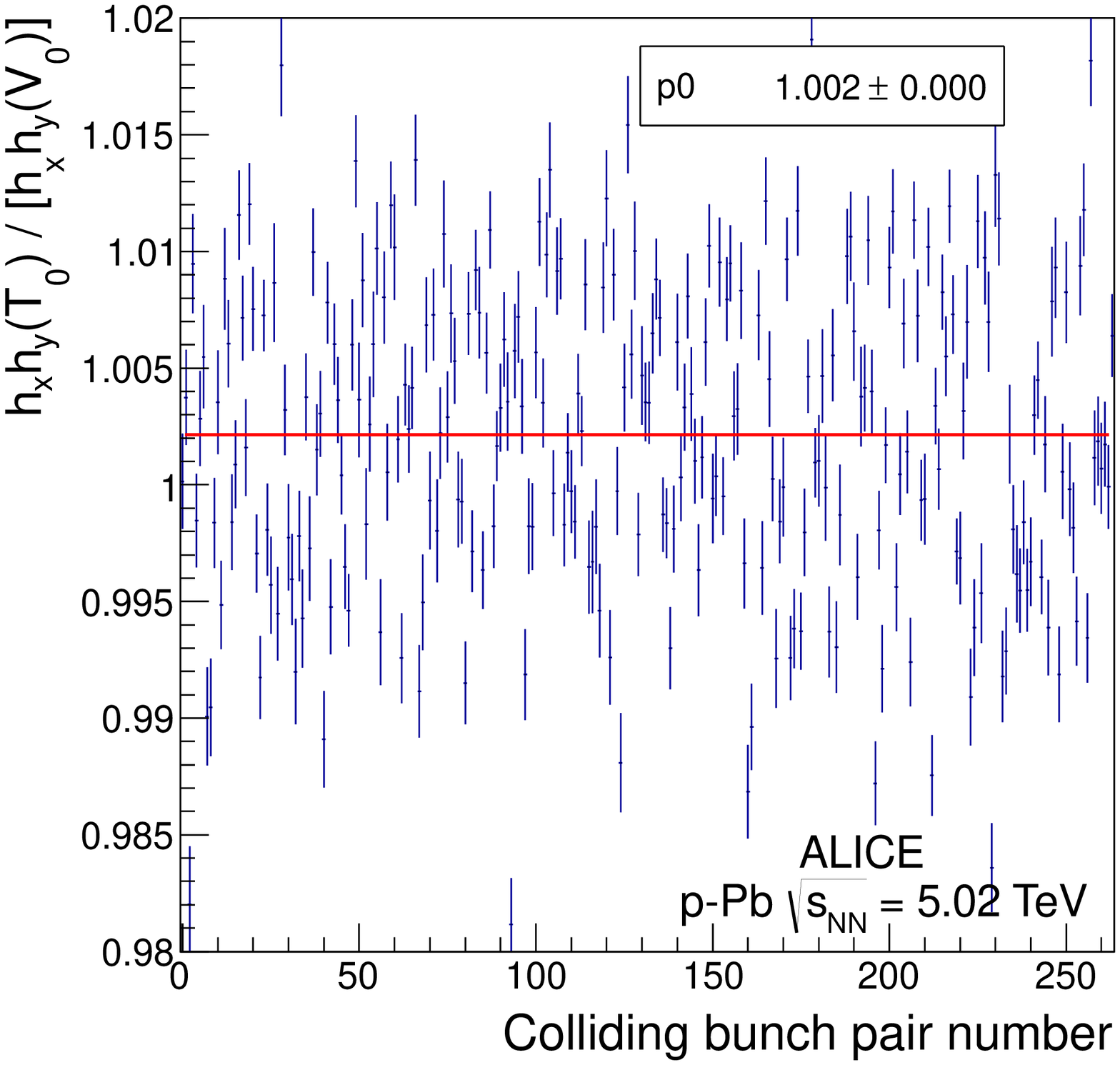} 
\includegraphics[width=0.49\textwidth, height=0.52\textwidth]{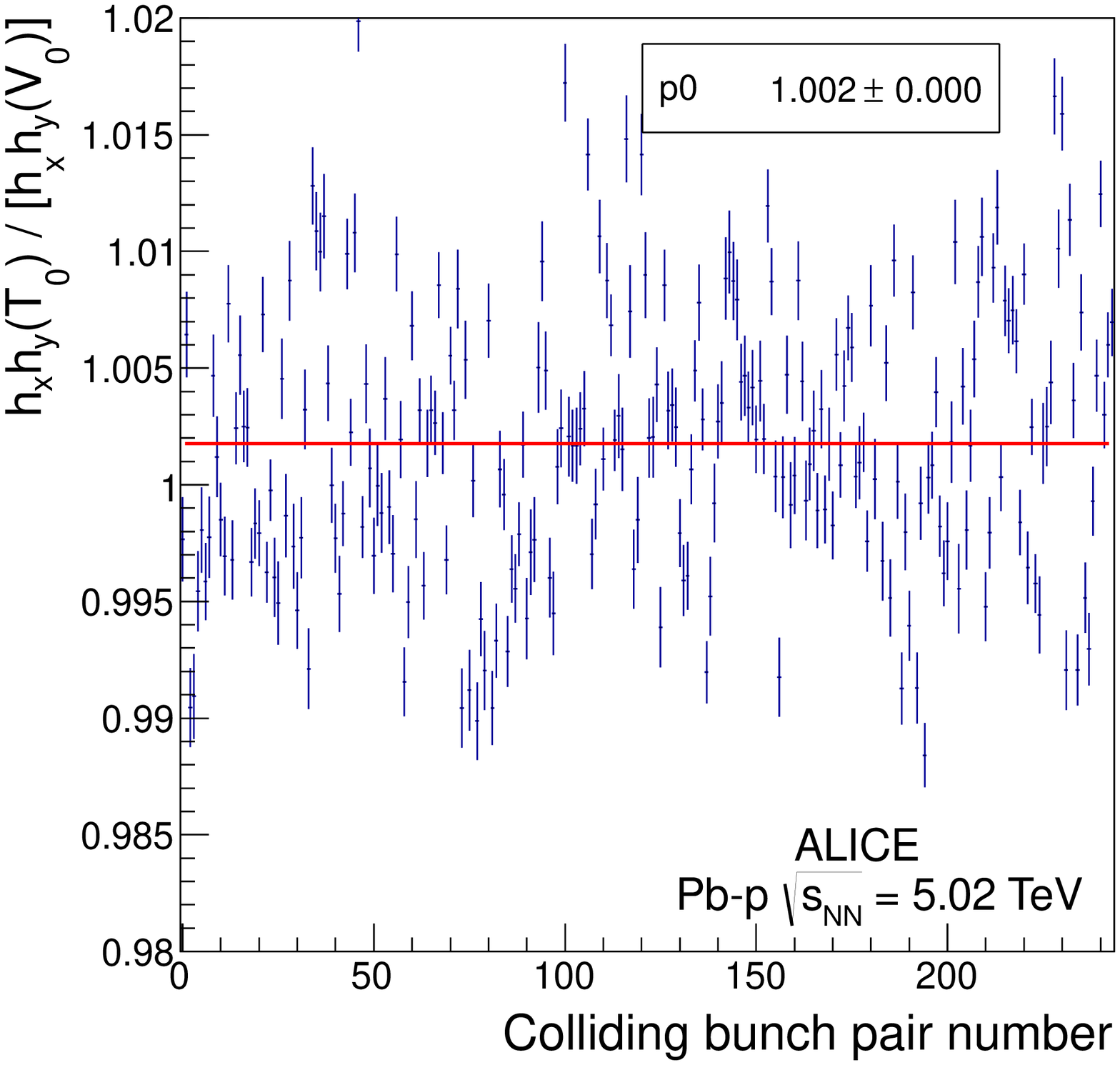} \\
\end{center}
   \caption{(Colour online)  Ratio between the $h_xh_y$ quantities obtained (via numerical method) with the T0 and V0 reference processes in the \hbox{p--Pb} (left) and \hbox{Pb--p} (right) scan session, as a function of the colliding bunch pair ID number. The solid red lines are zero-order-polynomial fits to the data.}
\label{fig:ShapeBunch}
\end{figure}

The corrected rates obtained with the above-described procedure are used to compute the effective beam widths $h_x$ and $h_y$. This is done with both a fit and a numerical method. 
For the fit method, it was found that a Gaussian (or double-Gaussian) function does not describe satisfactorily the measured shapes, while reasonable values of $\chi^2$ per degree of freedom ($\chi^2/ndf$~$\simeq$~1 on average, and typically $\chi^2/ndf$~$<$~2) are obtained by using a modified Gaussian function
\begin{equation}
R(\Delta x,0)  = R(0,0) \exp{[ -(\Delta x-\mu)^2/2\sigma^2]}\;[1+p_2(\Delta x-\mu)^2+ p_4(\Delta x-\mu)^4+p_6(\Delta x-\mu)^6]
\label{eq:fit}
\end{equation} 
and a similar one for \hbox{$R(0,\Delta y)$}.  An example of the quality of the fit is shown in figure \ref{fig:ShapeFit}. In the fit approach, the scan area  and the head-on rate $R(0,0)$ are obtained from the fit parameters.
In the numerical method, the scan area is obtained as the sum of all rates multiplied by the step size, and $R(0,0)$ is the measured rate at zero separation. 
The effective beam widths and head-on rates obtained with the two methods agree within 0.5\%. Since the effective beam widths are independent of the process used to measure them, a consistency check is performed by computing the ratio of the $h_xh_y$ quantities of equation~\ref{eq:Lumi}, obtained with  T0 and V0, for each colliding bunch. The results are shown in figure~\ref{fig:ShapeBunch}. For both scan sessions, the bunch-averaged value of the ratios is compatible with unity within 0.2\%.

\begin{figure}[tbp]
\begin{center}
\includegraphics[width=0.46\textwidth]{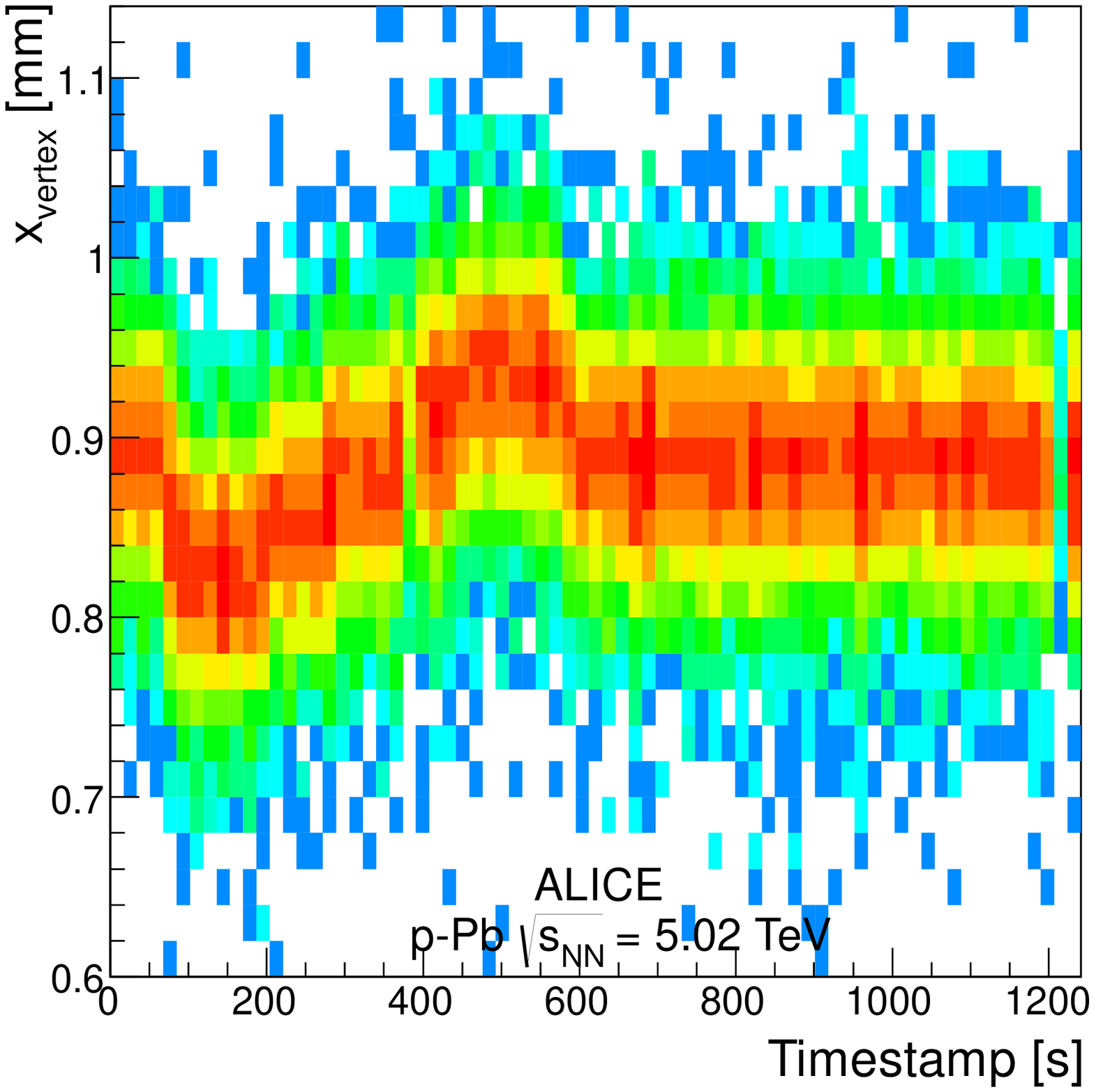} 
\includegraphics[width=0.48\textwidth]{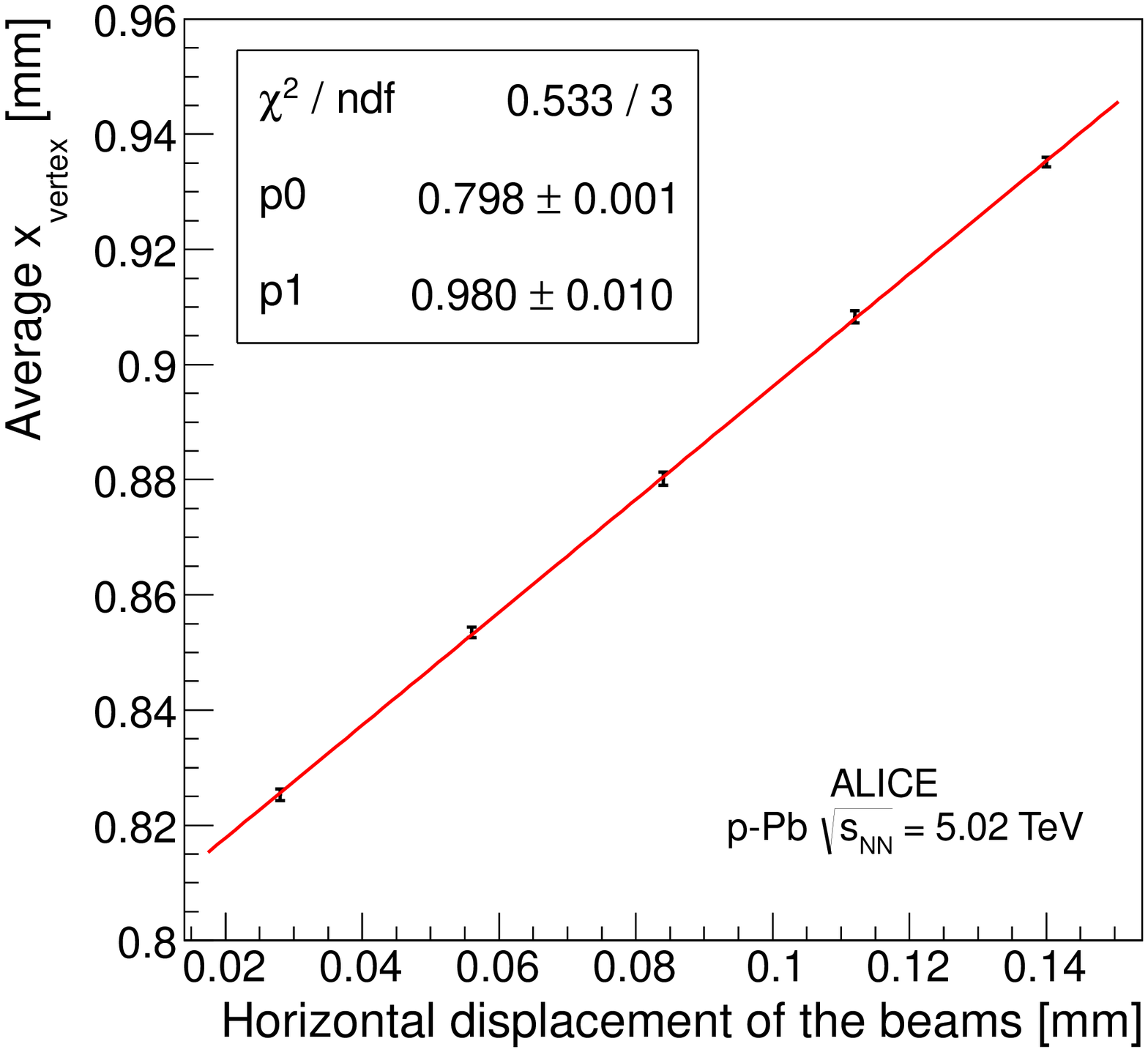} \\
\end{center}
   \caption{(Colour online) Left: distribution of the horizontal interaction vertex coordinate as a function of time during the length-scale calibration run. The structure visible in the timestamp region between $\simeq$100~s and $\simeq$600~s corresponds to the beams being moved in five steps of 28~$\mu$m each. Right: average horizontal vertex coordinate as a function of the nominal horizontal beam displacement in the length-scale calibration run, with superimposed linear fit (solid red line).}
\label{fig:lsc1}
\end{figure}

The measured beam widths are corrected by a length-scale calibration factor. This correction aims to fine tune the conversion factor (known with limited precision) between the current in the steering magnets and the beam displacement. The calibration is performed in a dedicated run, where the two beams are moved simultaneously in the same direction in steps of equal size; the changes in the interaction vertex position provide a measurement of the actual beam displacement, which is used to extract a correction factor to the nominal displacement scale. The displacement of the vertex position is measured using data from the ALICE Inner Tracking System~\cite{aliITS} and Time Projection Chamber~\cite{aliTPC}. This is shown in figure~\ref{fig:lsc1}, left, for the horizontal length-scale calibration run. For each step, the vertex position and its uncertainty are obtained from a Gaussian fit to the vertex distribution. The length-scale correction factor is obtained as the slope parameter of a linear fit to the measured vertex displacement as a function of the nominal displacement (figure~\ref{fig:lsc1}, right). Since this correction affects the global beam-displacement scale, all measured beam widths are multiplied by the correction factors 0.98$\pm$0.01 for the horizontal scale and  1.02$\pm$0.01 for the vertical scale.

\begin{figure}[tbp]
\begin{center}
\includegraphics[width=0.48\textwidth]{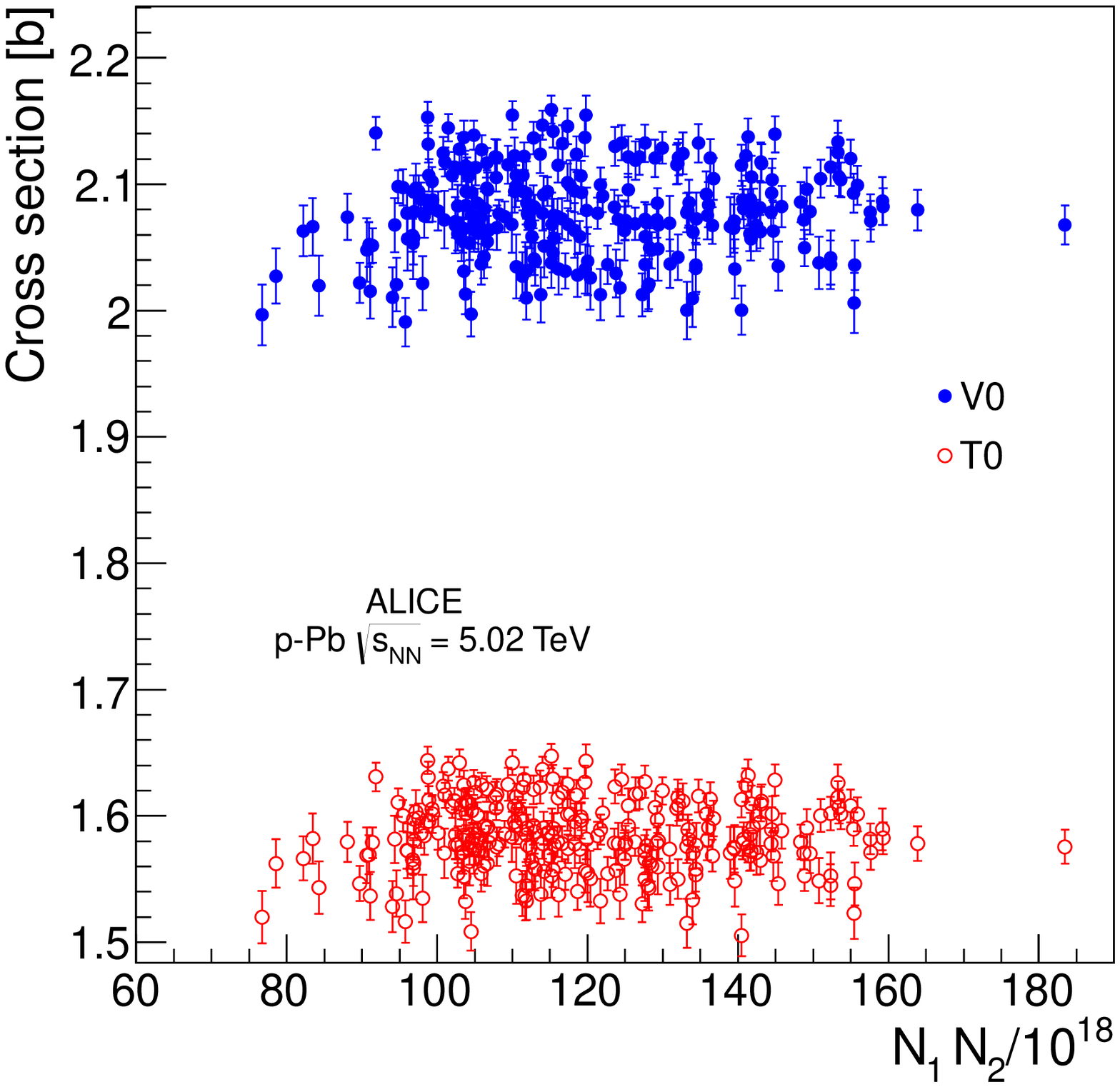} 
\includegraphics[width=0.48\textwidth]{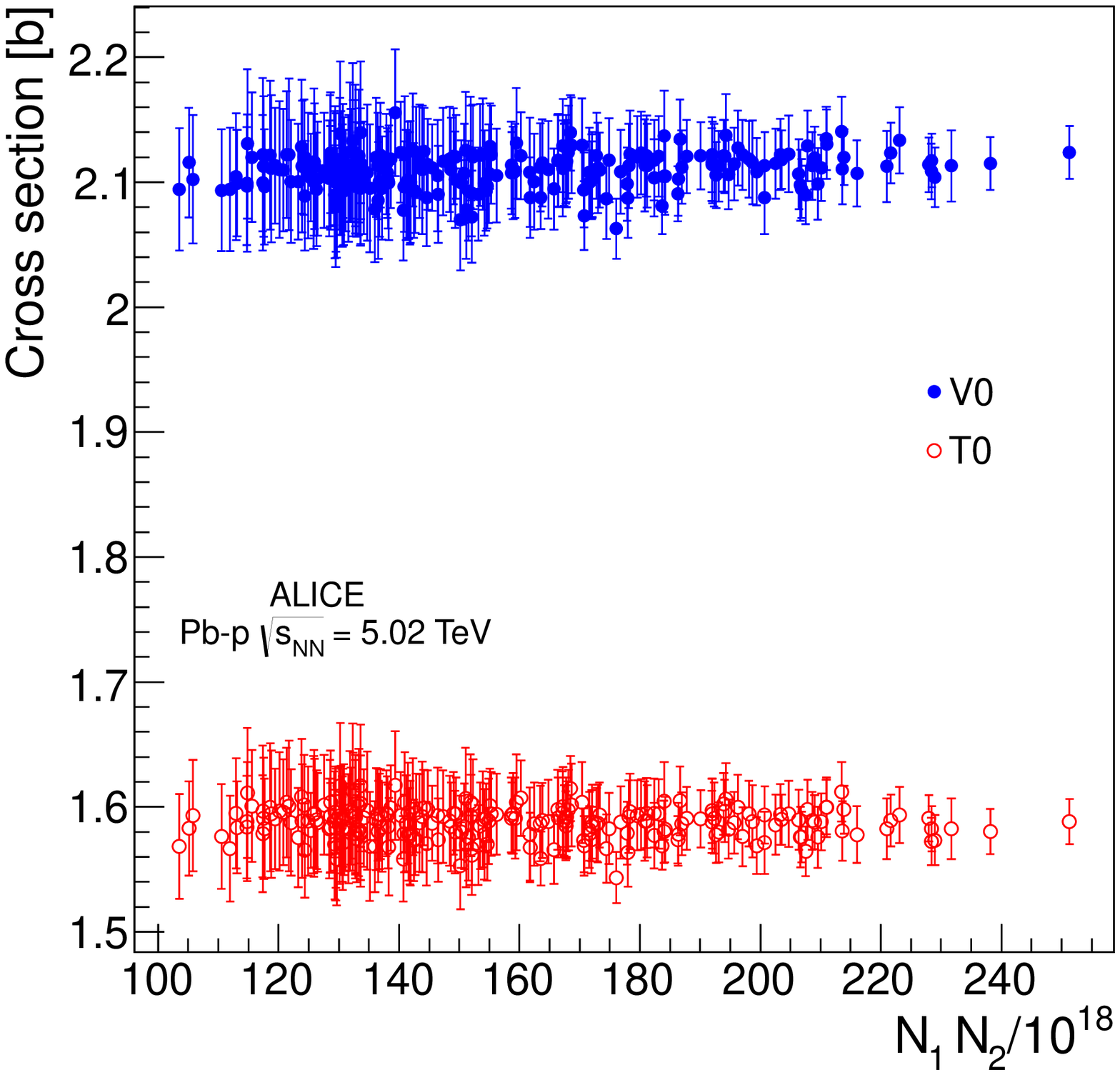} \\ 
\end{center}
   \caption{(Colour online) Cross sections for the T0 and V0 processes measured in the first scan of the p--Pb (left) and Pb--p (right) sessions,  as a function of the product of the intensities of the colliding bunch pair.  The results are obtained with the numerical method. Only the statistical uncertainties are shown.}
\label{fig:xSecNN}
\end{figure}

The cross section for each colliding bunch pair and reference process is calculated according to eq.~\ref{eq:Lumi} and \ref{eq:crSec}  from the measured bunch intensities, beam widths and head-on rates. As there are two measured head-on rates per scan pair (one from the vertical and one from the horizontal scan), the arithmetic mean of the two is used, after checking that the two values are compatible within statistical uncertainties.

The measured cross sections (obtained with the numerical method) for the T0- and V0-based processes during the first scan of the p--Pb and Pb--p sessions are shown in figure~\ref{fig:xSecNN} for all the colliding bunch pairs, as a function of the product of the colliding bunch intensities ($N_1N_2$). No dependence of the results on $N_1N_2$ is observed. For the p--Pb session, fluctuations beyond the statistical uncertainties are observed, and accounted for as a source of systematic uncertainty (see section~\ref{sec:results}).

\section{Results and systematic uncertainties}\label{sec:results}

For both processes and scan sessions the weighted average of results from all colliding bunch pairs is computed, for each scan and method. The results for all scans and methods are summarised in tables~\ref{tab:allResPpb} and \ref{tab:allResPbp}. The numerical and fit method agree to better than 0.3\% for all scans. The numerical result is preferred, because it implies a weaker assumption on the scan shape and to be consistent with earlier ALICE results in pp and Pb--Pb collisions~\cite{aliPerf,aliInel}. For each session, the weighted average of the results of the two performed scans is retained as the final result. The differences between the two methods and between different scans in the same session are taken into account in the evaluation of the systematic uncertainties.

\begin{table}[tbp] \small
\begin{tabular}{l|ccc|ccc} 
 Method &  $\sigma_{\rm{V0}}$ [b] & & & $\sigma_{\rm{T0}}$ [b] & &\\  \hline
        &  First scan & Second scan & Average & First scan & Second scan & Average\\  
Num.	& 2.087$\pm$0.001  & 2.098$\pm$0.001 & 2.093$\pm$0.001 & 1.590$\pm$0.001 & 1.598$\pm$0.001 & 1.594$\pm$0.001 \\  
Fit & 2.086$\pm$0.001  & 2.099$\pm$0.001 &  & 1.595$\pm$0.001& 1.602$\pm$0.001&\\  \hline
\end{tabular}
\caption{Cross section for the V0- and T0-based reference process in two p--Pb vdM scans, as obtained with the numerical and fit methods. The weighted average between the numerical results of the two scans, retained as the final result, is also reported. The quoted uncertainties are statistical. }
\label{tab:allResPpb}
%\vspace{0.2 cm}
\end{table}

\begin{table} \small
\begin{tabular}{l|ccc|ccc}  
 Method &  $\sigma_{\rm{V0}}$ [b] & & & $\sigma_{\rm{T0}}$ [b] & &\\  \hline
        &  First scan & Second scan & Average & First scan & Second scan & Average\\  
Num. 	& 2.110$\pm$0.002  & 2.141$\pm$0.003 & 2.122$\pm$0.002 & 1.586$\pm$0.002 & 1.607$\pm$0.003 & 1.594$\pm$0.002\\  
Fit &  2.105$\pm$0.002 & 2.138$\pm$0.002 &  & 1.581$\pm$0.002& 1.605$\pm$0.002 &\\  \hline
\end{tabular}
\caption{Cross section for the V0- and T0-based reference process in two Pb--p vdM scans, as obtained with the numerical and fit methods. The weighted average between the numerical results of the two scans, retained as the final result, is also reported. The quoted uncertainties are statistical.}
\label{tab:allResPbp}
%\vspace{0.2 cm}
\end{table}

\begin{figure}[tbp]
\begin{center}
\includegraphics[width=0.48\textwidth]{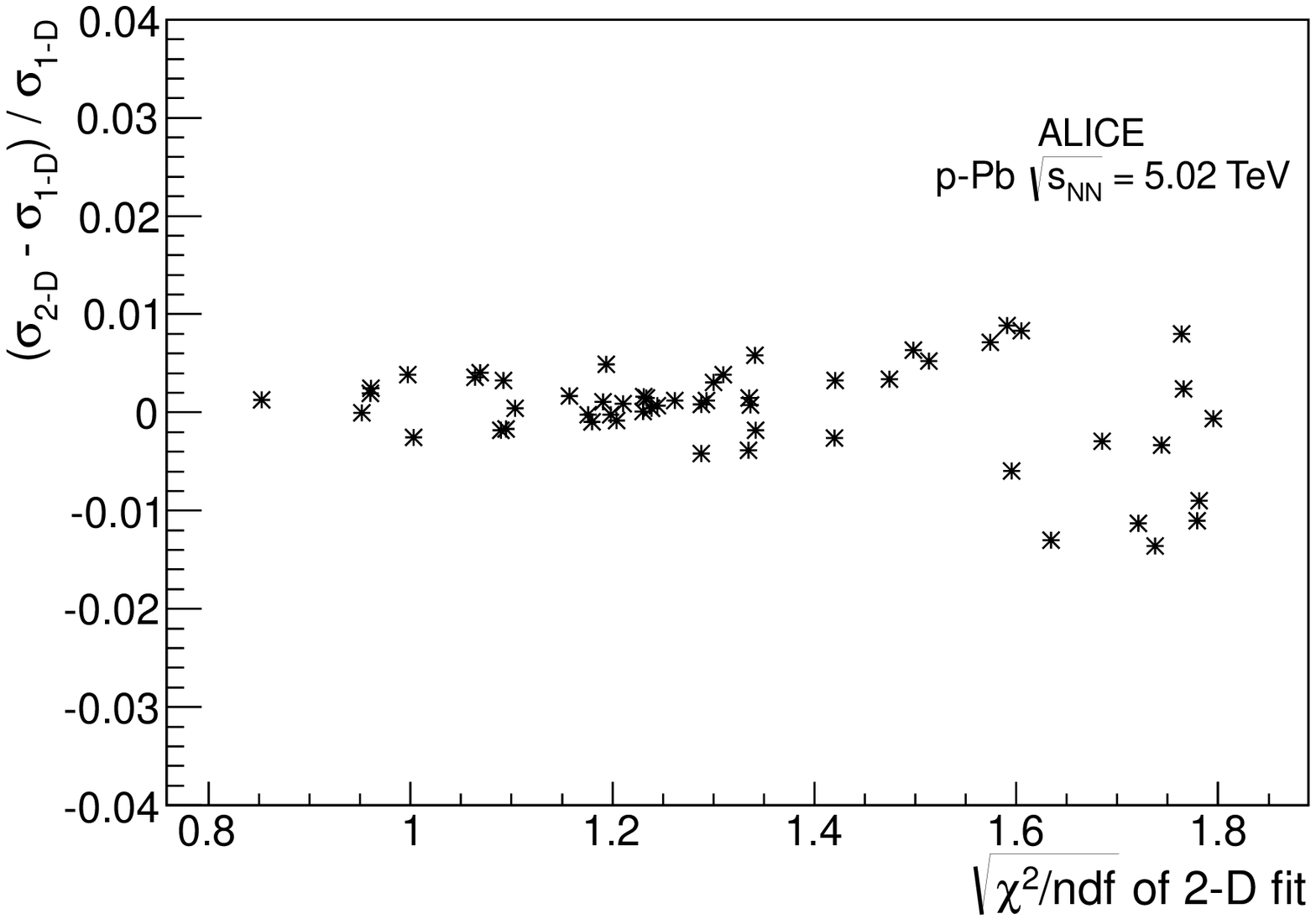} 
\includegraphics[width=0.48\textwidth]{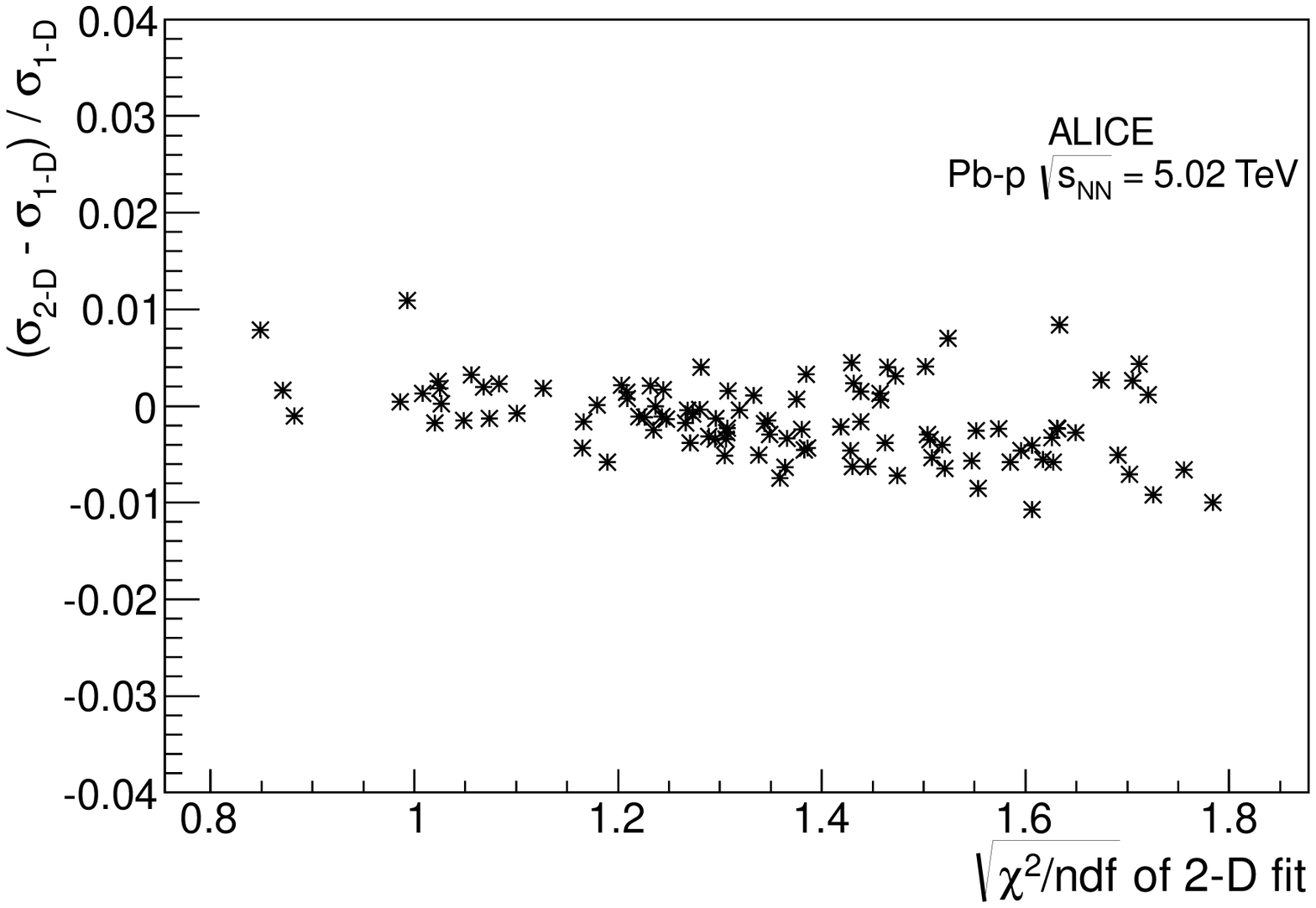} \\ 
\end{center}
   \caption{Relative difference between the cross section obtained with the two-dimensional DG2D and the one-dimensional gPol6 fit models, as a function of the $\sqrt{\chi^{2}/ndf}$ value of the two-dimensional fit, for $\chi^{2}/ndf$~$<$~105/32. Left: results for V0 in the first p--Pb scan. Right: results for V0 in the first Pb--p scan. }
\label{fig:TrCorr}
\end{figure}

The sources of systematic uncertainty considered are listed below; unless otherwise specified, the quoted uncertainties apply to both the T0 and the V0 cross-section measurements.

\begin{itemize}
\item  Transverse correlations: the formalism of equation~\ref{eq:Lumi} assumes complete factorisation of the beam profiles in the two transverse directions, such that the beam overlap region is fully described by the $h_x h_y$ quantity. Luminosity measurements at the LHC~\cite{c_barschel_phd,cmsLumi1} have shown that factorisation can be broken at a non-negligible level. In previous studies~\cite{cmsLumi1,atlasLumi,cmsLumi2}, the bias arising from such an effect was quantified by comparing the results obtained with the standard analysis method with those obtained from a correlated two-dimensional double-Gaussian fit to the data.  The same method was applied to this analysis, using the fitting function defined  in~\cite{cmsLumi2}. As already observed for the one-dimensional case, the double-Gaussian function provides a rather poor description of the data, with $\chi^2/ndf$ ranging from~$\simeq$~140/32 to~$\simeq$~1200/32  across scans and bunches. A better agreement ($\chi^2/ndf$ ranging from~$\simeq$~20/32 to~$\simeq$~800/32) is found if the standard double-Gaussian function is modified by dropping the requirement  that the coefficients of both Gaussian functions be positive (unconstrained double-Gaussian). The bunch-by-bunch difference between the cross section obtained with the unconstrained two-dimensional double-Gaussian (DG2D) fit model and that obtained with the one-dimensional fit model of Eq.~\ref{eq:fit} (gPol6)  shows a two-fold behaviour, depending on how well the scan shape is reproduced by the DG2D fit. For fits with relatively small $\chi^2/ndf$, the difference fluctuates by at most $\simeq$~1\% around zero and shows no dependence on $\chi^2/ndf$. For fits with large $\chi^2/ndf$, the difference is large (up to~$\simeq$~5\%) and systematically negative, and it exhibits a strong dependence on $\chi^2/ndf$, its magnitude increasing linearly with $\sqrt{\chi^2/ndf}$.  The negative difference values observed in the high-$\chi^2/ndf$ region, as well as their trend as a function of $\chi^2/ndf$, are identically observed when comparing the results of a one-dimensional double-Gaussian (DG1D) fit with those of the gPol6 fit. Hence, they are not interpreted as the effect of transverse correlations, but rather as a fit-model bias occurring when the scan shape is not well reproduced by the double-Gaussian fit. Such an interpretation is supported by the previously-cited studies~\cite{c_barschel_phd,cmsLumi1, atlasLumi}, where it is reported that the presence of transverse correlation systematically leads to positive discrepancies between the two-dimensional and the one-dimensional fit. %In the small-$\chi^2/ndf$ region, the DG2D result is in general slightly larger than the DG1D result, which points to a genuine (although small) transverse correlation effect. 
For bunches in the small-$\chi^2/ndf$ region, where no fit bias is expected, the transverse correlation uncertainty is evaluated as the full envelope of the bunch-by-bunch difference between the DG2D and the gPo6 fit result. Operationally, the small-$\chi^2/ndf$ region is defined as the one where there is no correlation between the DG2D-gPol6 difference and the value of $\chi^2/ndf$. The threshold $\chi^2/ndf$ value corresponding to such a definition varies slightly across scans and luminometers,  ranging from 58/32 to 105/32. For both the p--Pb and Pb--p scan sessions, the full envelope is at most 2.2\% across scans and luminometers (Fig.~\ref{fig:TrCorr}), hence such a value is retained as uncertainty.  For bunches in the large-$\chi^2/ndf$ region the uncertainty could not be evaluated with the above-described method, due to the double-Gaussian fit bias. If these bunches are affected by transverse correlations in a different way than the small-$\chi^2/ndf$ bunches, they may in principle bias the measured, bunch-averaged, cross section. In order to quantify a possible bias, the average cross sections obtained from the two sub-sets of bunches are compared. They are found to be consistent within 1.3\% (0.6\%) for the p--Pb (Pb--p) scan session. These values are assigned as additional uncertainty, leading to a total transverse-correlation uncertainty of 2.6\% (2.3\%) for the p--Pb (Pb--p) cross sections. 
\item Bunch-by-bunch consistency: an uncertainty of 1.6\% is assigned to the results of the \hbox{p--Pb} session. It is obtained from the RMS of the distribution of the cross section measured for all colliding bunch pairs, after subtracting in quadrature the bunch-averaged statistical uncertainty. For the Pb--p session the RMS is smaller than the average statistical uncertainty, hence no systematic uncertainty is assigned.
\item Scan-to-scan consistency: the difference between the first and second scan in the same session  (0.5\% for the p--Pb scans and 1.5\% for the Pb--p scans)  is retained as a systematic uncertainty.  
\item Length-scale calibration: 1.5\%, from the quadratic sum of the statistical uncertainties on the horizontal and vertical scale factors reported in section~\ref{sec:analysis}.
\item Background subtraction: in order to evaluate a possible bias arising from beam-beam events identified as beam-gas by the cut described in section~\ref{sec:analysis}, the analysis has been repeated by increasing the width of the window for beam-beam events from 8 to 14~ns: for the V0 cross section, a difference of 0.45\% is found and added to the systematic uncertainty for both configurations. The difference is negligible ($\ll$~0.1\%) for the T0 cross section.
\item Method dependence: 0.3\% for both scan sessions, quantified via the maximum difference between the results obtained with the numerical and the fit method (tables~\ref{tab:allResPpb} and \ref{tab:allResPbp}).
\item Beam centering: the measurement of $R(0,0)$ can be affected by a non-optimal alignment of the two beams in the head-on position. Such misalignment is quantified, for the $x$ and $y$ directions, via the $\mu$ parameter of eq.~\ref{eq:fit}. For about half of the scans, the value of $\mu$ is compatible with zero; for the first horizontal and the second vertical p-Pb scan, and for the first horizontal Pb-p scan, it reaches up to 2.5~$\mu$m. The effect of such misalignment on the measured head-on rates was estimated using eq.~\ref{eq:fit} and the obtained fit parameters: the resulting systematic uncertainty on the cross-section measurement is 0.3\% (0.2\%) for the p--Pb (Pb--p) configuration.
\item Trigger dependence of the measured beam widths: 0.2\% for both sessions, from the bunch-averaged difference between the $h_xh_y$ quantities measured with T0 and V0 (figure~\ref{fig:ShapeBunch}).
\item Luminosity-decay correction: when varying the luminosity decay parameters within their uncertainties, a negligible ($<$~0.1\%) effect on the measured cross section is observed.
\item Bunch intensity: the uncertainty on the bunch-intensity product $N_1N_2$ arising from the DCCT calibration~\cite{dcct_note} is 0.46\% (0.54\%) for the p--Pb (Pb--p) scan session; given the very large fraction of colliding over circulating bunches, the uncertainty on the relative bunch populations has negligible effect on the cross section measurement~\cite{bcnwg_relpop}.
\item Orbit drift: possible variations of the reference orbit during the scan may lead to a difference between the nominal and the real beam separation. In order to quantify a possible bias, the data from the LHC Beam Position Monitors (BPM)~\cite{bpm} in various locations along the ring are used  to extrapolate, with the YASP steering program~\cite{yasp}, the transverse coordinates of the reference orbit of the two beams at IP2, for each scan step. The (small) observed variations in the orbit are used to correct the separation values, and the cross section is re-calculated: a difference of 0.4\% (0.1\%) is found for the p--Pb (Pb--p) configuration results. 
\item Beam-beam deflection: due to their electric charge, the two beams exert a repulsive force upon each other~\cite{kickNote}. Such repulsion results in a beam separation slightly different than its nominal value. The variations of the beam separation are calculated using the MAD-X~\cite{madx} code: the effect on the measured cross section (partially correlated between the p--Pb and Pb--p sessions) is found to be 0.2\% (0.3\%) for the p--Pb (Pb--p) scan, in the same direction for the two fills. 
\item Ghost and satellite charge: the uncertainty on the LHCb ghost-charge measurement~\cite{c_barschel_phd} propagates to an uncertainty of 0.1\% (0.2\%) on the p--Pb (Pb--p) cross-section measurement; the uncertainty in the LDM satellite-charge measurement~\cite{ldm_note} propagates to an uncertainty of 0.04\% (0.1\%) on the p--Pb (Pb--p) cross-section measurement.
\item Dynamic $\beta^*$: due to their electric charge, the two colliding beams (de-)focus each other in a separation-dependent way, which alters the measured scan shape. Calculations~\cite{wHerr} are used to estimate the variations of $\beta^*$ with the separation, according to the prescription given in~\cite{atlasLumi}; the effect on the measured cross section (partially correlated between the p--Pb and Pb--p sessions)  is found to be $\leq$~0.1\%  for all p--Pb and Pb--p scans.
\end{itemize} 

Summing in quadrature all the above-mentioned uncertainties (summarised in table~\ref{tab:uncertainties}), one gets a total systematic uncertainty of 3.5\% for the p--Pb cross sections and 3.2\% for the Pb--p cross sections. The uncertainty applies in the same way to the T0 and V0 cross sections, since the only non-common term is the background subtraction, which becomes negligible in the quadratic sum.

The final results for the p--Pb configuration are
\begin{displaymath}
%\sigma_{\rm{V0}} = 2.09\:\rm{b}\pm3.5\% = (2.09\pm0.07)\:\rm{b}\:\rm{,} \;\;\;  \sigma_{\rm{T0}} = 1.59\:\rm{b}\pm3.5\% = (1.59\pm0.06)\:\rm{b} 
\sigma_{\rm{V0}} = (2.09\pm0.07)\:\rm{b}\:\rm{,} \;\;\;  \sigma_{\rm{T0}} = (1.59\pm0.06)\:\rm{b} 
\end{displaymath}
 and those for the Pb--p configuration are
\begin{displaymath}
%\sigma_{\rm{V0}} = 2.12\:\rm{b}\pm3.2\% = (2.12\pm0.07)\:\rm{b}\:\rm{,} \;\;\;  \sigma_{\rm{T0}} = 1.59\:\rm{b}\pm3.2\% = (1.59\pm0.05)\:\rm{b}\rm{.}
\sigma_{\rm{V0}} = (2.12\pm0.07)\:\rm{b}\:\rm{,} \;\;\;  \sigma_{\rm{T0}}  = (1.59\pm0.05)\:\rm{b}\rm{.}
\end{displaymath}
All uncertainties are systematic.

The length-scale calibration and background-subtraction uncertainties are fully correlated between the p--Pb and Pb--p results,  leading to a total correlated uncertainty between the two measurements of 1.5\% for T0 and 1.6\% for V0.

%The measured V0 cross section for Pb--p collisions is compatible, within uncertainties, with the visible cross section of (2.09$\pm$0.12)~b measured by the LHCb experiment for an equivalent beam configuration in a similar pseudo-rapidity range (\hbox{$3<\eta<5$})~\cite{LHCb}.

\begin{table} \small
\begin{tabular}{l|cc|c} 
 Uncertainty & p--Pp & Pb--p & Correlated between p--Pb and Pb--p \\ \hline
 Transverse correlations & 2.6\% & 2.3\% & No\\
 Bunch-by-bunch consistency & 1.6\% & - & No \\
 Scan-to-scan consistency & 0.5\% & 1.5\% & No\\
 Length-scale calibration & 1.5\% & 1.5\%& Yes \\
 Background subtraction (V0 only) & 0.5\% & 0.5\% & Yes \\
 Method dependence & 0.3\% & 0.3\%& No\\ 
 Beam centering    & 0.3\% & 0.2\%& No \\ 
 Bunch size vs trigger & 0.2\% & 0.2\% & No \\
 Bunch intensity   & 0.5\% & 0.5\% & No  \\
 Orbit drift       & 0.4\% & 0.1\% & No \\
 Beam-beam deflection &  0.2\%    &  0.3\%    &  Partially \\ 
 Ghost charge      &0.1\% & 0.2\%& No  \\
 Satellite charge  &$<$0.1\% & 0.1\%& No  \\
 Dynamic $\beta^*$ &    $<$0.1\%     &  0.1\%    &  Partially   \\ \hline
 
 Total on visible cross section & 3.5\% & 3.2\% & \\ \hline
 V0- vs T0-based integrated luminosity & 1\% & 1\% & No \\ \hline
 Total on integrated luminosity & 3.7\% & 3.4\% & \\ \hline
\end{tabular}
\caption{Relative uncertainties on the measurement of the T0 and V0 reference process cross section in p--Pb and Pb--p collisions.}
\label{tab:uncertainties}
%\vspace{0.2 cm}
\end{table}

\section{Comparison between V0- and T0-based luminosities} \label{comparison}

The visible cross sections measured in the vdM scans are used to determine the integrated luminosity for the data collected in the 2013 proton-lead run~\cite{jpsi_pPb,D_pPb,psi2s_pPb,UPC_pPb}. The luminosity is measured independently via the V0 or the T0 trigger counts, corrected for pileup and for background contamination in the same way as done for the vdM scan data, divided by the corresponding cross sections.

\begin{figure}[tbp]
\begin{center}
\includegraphics[width=0.48\textwidth]{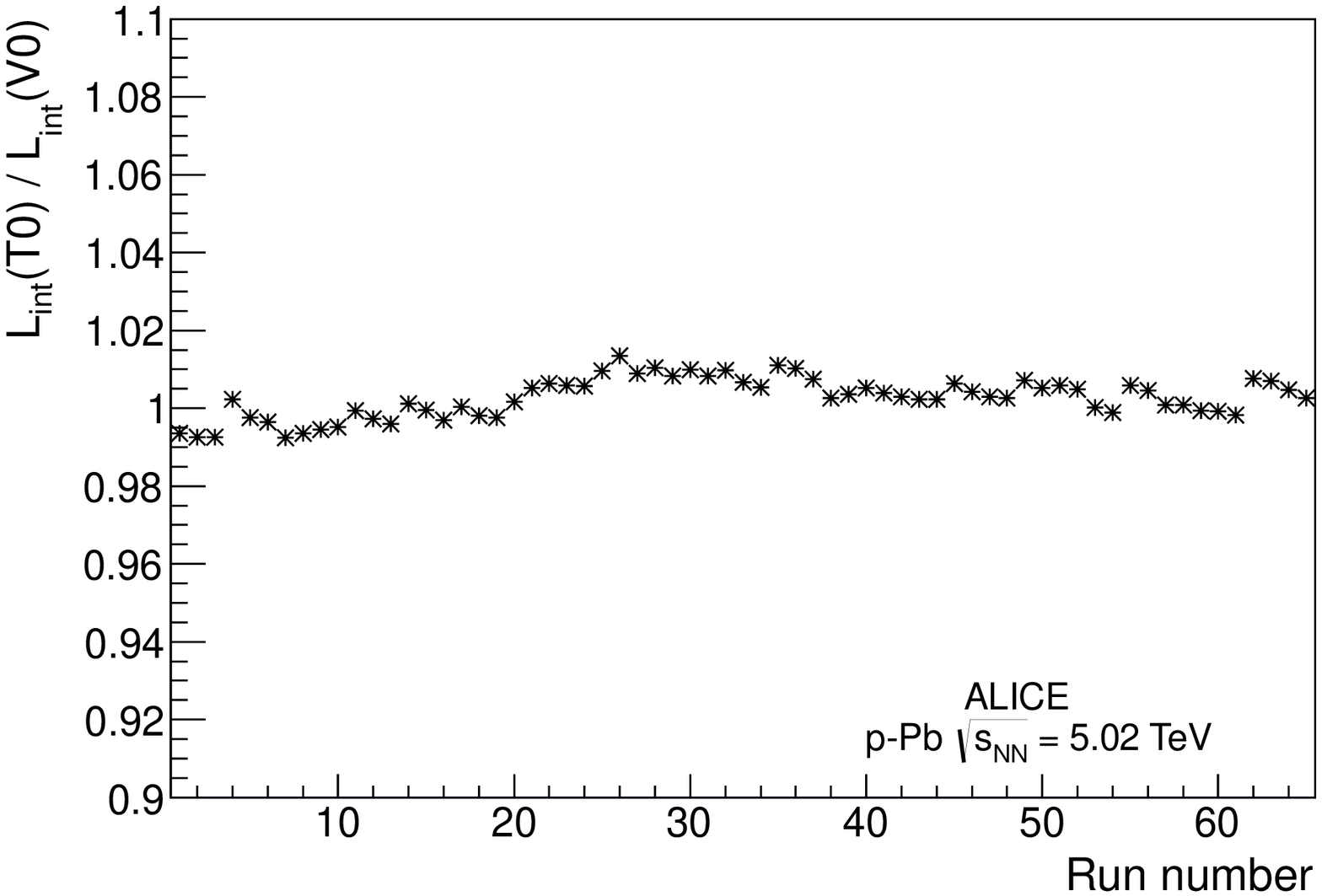} 
\includegraphics[width=0.48\textwidth]{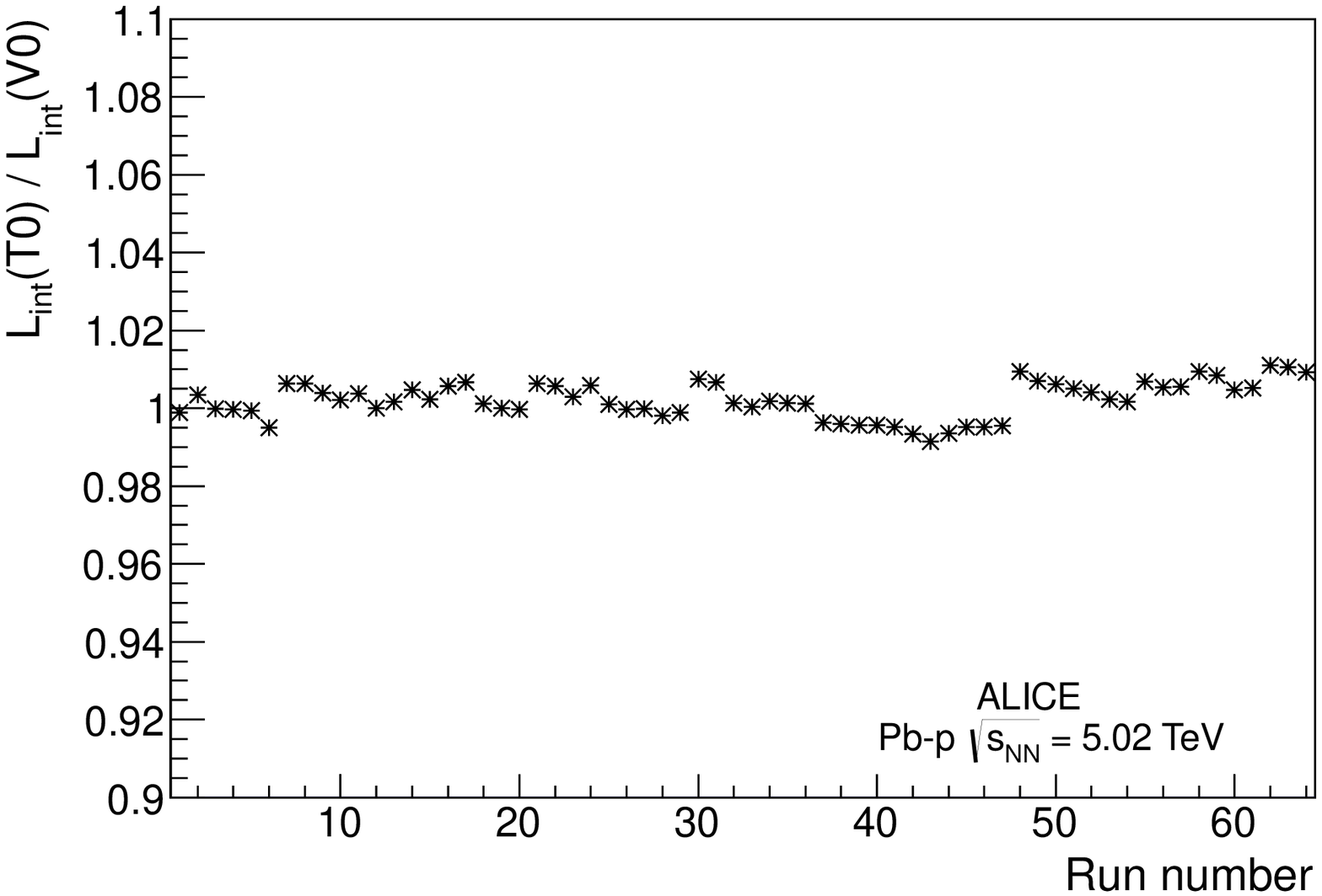} \\
\end{center}
   \caption{Ratio of T0- to V0-based integrated luminosities as a function of run number for the p--Pb (left) and Pb--p (right) data taking. The tiny statistical uncertainties are covered by the data-point markers.}
\label{fig:lumiRatio}
\end{figure}

The data sample is divided in several  smaller datasets (runs). The integrated luminosity corresponding to each run is computed using both reference processes, and the results are compared. Figure~\ref{fig:lumiRatio}  shows the ratio of the integrated luminosity obtained with T0 to the one obtained with V0, as a function of the run number, for the p--Pb and Pb--p running modes. Although the overall agreement is satisfactory, fluctuations of about 1\% around unity are seen in the run-by-run ratio; since these fluctuations are significantly larger than the tiny statistical uncertainties, a 1\% additional systematic uncertainty is considered in the computation of the integrated luminosity uncertainty (table~\ref{tab:uncertainties}).

\section{Measurement of the ZDC trigger cross section} \label{zdc}

The ALICE Zero Degree Calorimeter system (ZDC)~\cite{aliZDC}  is composed of two neutron (ZN) and two proton (ZP) calorimeters, as well as two small electromagnetic calorimeters (ZEM). 
The two ZNs (ZNA and ZNC) are located on opposite sides of IP2, 112.5 m away from the interaction point. Each ZN is placed at zero degrees with respect to the ALICE $z$ axis
and is used to detect neutral particles at pseudo-rapidities $| \eta |  >$~8.8. 
The ZNs were used to measure the cross section for neutron emission in Pb--Pb collisions at the LHC~\cite{emd}.  A similar study is foreseen in p--Pb collisions. For this purpose, data have been collected with a trigger condition requiring a signal in the ZN located on the Pb remnant side (i.e. ZNA for p--Pb, ZNC for Pb--p). In this paragraph,  the measured T0 and V0 cross sections are used to determine indirectly the cross section for events satisfying such a trigger condition. Since the trigger condition is symmetric with respect to the swapping of the proton and lead beams, one expects the cross section to be the same in the p--Pb and Pb--p configurations. Thus, such a measurement provides a consistency check for the analysis of data from the two sessions.

The ZDC trigger cross section is calculated from the measured T0 and V0 cross sections, rescaled by the ratio of the ZDC trigger rate to the rate of the two reference processes, as measured during the two vdM scan sessions. All rates are corrected for background and pileup. The ratios and the resulting cross sections for the ZDC trigger are reported in table~\ref{tab:zdcRatios}. The results obtained in the two fills are compatible within the uncorrelated uncertainties. The results obtained with T0 and V0 are also compatible. Thus, all results are combined to get

\begin{displaymath}
%\sigma_{\rm{ZDC}} = 2.22\:\rm{b}\: \pm0.3\%\:(stat)\:\pm 2.5\%\:(syst) = 2.22\:\rm{b}\pm0.01\:\rm{b}\:(stat)\:\pm0.06\:\rm{b}\:(syst)\rm{.}
\sigma_{\rm{ZDC}} =  2.22\:\rm{b}\pm0.01\:\rm{b}\:(stat)\:\pm0.06\:\rm{b}\:(syst)\rm{.}
\end{displaymath}

\begin{table} \small
\begin{tabular}{lccc}  
 Configuration & Reference & $R_{\rm{ZDC}}$/$R_{\rm{Reference}}$ & $\sigma_{\rm{ZDC}}$~=~$\sigma_{\rm{Reference}}\frac{R_{\rm{ZDC}}}{R_{\rm{Reference}}}$ [b] \\ \hline
 p--Pb & T0 & 1.380$\pm$0.014 (stat) & 2.20 $\pm$ 0.02 (stat) $\pm$ 0.07 (uncorr.) $\pm$ 0.03 (corr.) \\
 p--Pb & V0 & 1.046$\pm$0.012 (stat) & 2.19 $\pm$ 0.02 (stat) $\pm$ 0.07 (uncorr.) $\pm$ 0.03 (corr.) \\
 Pb--p & T0 & 1.404$\pm$0.005 (stat) & 2.24 $\pm$ 0.01 (stat) $\pm$ 0.06 (uncorr.) $\pm$ 0.03 (corr.) \\
 Pb--p & V0 & 1.050$\pm$0.004 (stat) & 2.23 $\pm$ 0.01 (stat) $\pm$ 0.06 (uncorr.) $\pm$ 0.03 (corr.) \\\hline
\end{tabular}
\caption{Ratio of the ZDC to the reference process rates and ZDC cross sections resulting from such ratios, for all reference processes and beam configurations. The systematic uncertainties are split into correlated and uncorrelated components between the p--Pb and the Pb--p sessions.}
\label{tab:zdcRatios}
%\vspace{0.2 cm}
\end{table}

\section{Conclusions} Van der Meer scans were done for proton-lead collisions at $\sqrt{s_{\rm{NN}}}$~=~5.02~TeV at the LHC. The cross section was measured for two reference processes, based on particle detection by the T0 (\hbox{$4.6<\eta< 4.9$} and \hbox{$-3.3<\eta<-3.0$}) and V0  (\hbox{$2.8<\eta< 5.1$} and \hbox{$-3.7<\eta<-1.7$}) detectors. For the p--Pb configuration (proton beam travelling clockwise), the cross-section uncertainty is 3.5\% and the results are: $\sigma_{\rm{V0}}$~=~2.09~b~$\pm$~0.07~b~(syst), $\sigma_{\rm{T0}}$~=~1.59~b~$\pm$~0.06~b~(syst). For the Pb--p configuration (proton beam travelling counter-clockwise), the cross-section uncertainty is 3.2\% and the results are: $\sigma_{\rm{V0}}$~=~2.12~b~$\pm$~0.07~b (syst), $\sigma_{\rm{T0}}$~=~1.59~b~$\pm$~0.05~b~(syst). The two reference processes were independently used for the luminosity determination in the 2013 proton-lead run at the LHC. The luminosities measured via the two processes  differ by at most 1\% throughout the whole data-taking period; with such value quadratically added to the reference process cross section uncertainties, a total uncertainty on the luminosity measurement of 3.7\% (3.4\%) for the \hbox{p--Pb} \hbox{(Pb--p)} configuration is obtained. Finally, the measured reference cross sections were used to indirectly determine the cross section for a third, configuration-independent, reference process, based on neutron detection by the Zero Degree Calorimeter: $\sigma_{\rm{ZDC}}$~=~2.22~b~$\pm$~0.01~b~(stat)~$\pm$~0.06~b (syst).

\newenvironment{acknowledgement}{\relax}{\relax}
\begin{acknowledgement}
\section{Acknowledgements}
\input{acknowledgements.tex}    %%%%%%% get the latest version before submitting
\end{acknowledgement}

\newpage
%
%\input{}               %%%%%%%%%%% put your appendices here
%
%%%%%%%%% appendix with author list
\appendix
\section{The ALICE Collaboration}
\label{app:collab}
\input{Alice_Authorlist_2014-Mar-11-CERNPREP.tex }  %%%%%%% get the latest version before submitting
\end{document}

%% file: acknowledgements.tex
The ALICE Collaboration would like to thank all its engineers and technicians for their invaluable contributions to the construction of the experiment; the CERN accelerator teams for the outstanding performance of the LHC complex, for availability and excellent support in carrying out vdM scans and for providing the bunch intensity and other crucial measurements; the LHCb collaboration, for providing the ghost charge measurements; everyone in the LHC Luminosity Calibration and Monitoring Working group, for many useful and stimulating discussions.
%\\
The ALICE Collaboration gratefully acknowledges the resources and support provided by all Grid centres and the Worldwide LHC Computing Grid (WLCG) collaboration.
%\\
The ALICE Collaboration acknowledges the following funding agencies for their support in building and
running the ALICE detector:
 %\\
State Committee of Science,  World Federation of Scientists (WFS)
and Swiss Fonds Kidagan, Armenia,
 %\\
Conselho Nacional de Desenvolvimento Cient\'{\i}fico e Tecnol\'{o}gico (CNPq), Financiadora de Estudos e Projetos (FINEP),
Funda\c{c}\~{a}o de Amparo \`{a} Pesquisa do Estado de S\~{a}o Paulo (FAPESP);
 %\\
National Natural Science Foundation of China (NSFC), the Chinese Ministry of Education (CMOE)
and the Ministry of Science and Technology of China (MSTC);
 %\\
Ministry of Education and Youth of the Czech Republic;
 %\\
Danish Natural Science Research Council, the Carlsberg Foundation and the Danish National Research Foundation;
 %\\
The European Research Council under the European Community's Seventh Framework Programme;
 %\\
Helsinki Institute of Physics and the Academy of Finland;
 %\\
French CNRS-IN2P3, the `Region Pays de Loire', `Region Alsace', `Region Auvergne' and CEA, France;
 %\\
German BMBF and the Helmholtz Association;
%\\
General Secretariat for Research and Technology, Ministry of
Development, Greece;
%\\
Hungarian OTKA and National Office for Research and Technology (NKTH);
 %\\
Department of Atomic Energy and Department of Science and Technology of the Government of India;
 %\\
Istituto Nazionale di Fisica Nucleare (INFN) and Centro Fermi -
Museo Storico della Fisica e Centro Studi e Ricerche "Enrico
Fermi", Italy;
 %\\
The Ministry of Education, Culture, Sports, Science and Technology (MEXT) and Japan Society for the Promotion of Science (JSPS), Japan
 %\\
Joint Institute for Nuclear Research, Dubna;
 %\\
%Korea Foundation for International Cooperation of Science and Technology (KICOS);
National Research Foundation of Korea (NRF);
 %\\
CONACYT, DGAPA, M\'{e}xico, ALFA-EC and the EPLANET Program
(European Particle Physics Latin American Network)
 %\\
Stichting voor Fundamenteel Onderzoek der Materie (FOM) and the Nederlandse Organisatie voor Wetenschappelijk Onderzoek (NWO), Netherlands;
 %\\
Research Council of Norway (NFR);
 %\\
Polish Ministry of Science and Higher Education;
 %\\
National Science Centre, Poland;
 %\\
 Ministry of National Education/Institute for Atomic Physics and CNCS-UEFISCDI - Romania;
 %\\
Ministry of Education and Science of Russian Federation, Russian
Academy of Sciences, Russian Federal Agency of Atomic Energy,
Russian Federal Agency for Science and Innovations and The Russian
Foundation for Basic Research;
 %\\
Ministry of Education of Slovakia;
 %\\
Department of Science and Technology, South Africa;
 %\\
CIEMAT, EELA, Ministerio de Econom\'{i}a y Competitividad (MINECO) of Spain, Xunta de Galicia (Conseller\'{\i}a de Educaci\'{o}n),
CEA\-DEN, Cubaenerg\'{\i}a, Cuba, and IAEA (International Atomic Energy Agency);
 %\\
Swedish Research Council (VR) and Knut $\&$ Alice Wallenberg
Foundation (KAW);
 %\\
Ukraine Ministry of Education and Science;
 %\\
United Kingdom Science and Technology Facilities Council (STFC);
 %\\
The United States Department of Energy, the United States National
Science Foundation, the State of Texas, and the State of Ohio.

%% file: Alice_Authorlist_2014-Mar-11-CERNPREP.tex
% Collaboration: CERN-LHC-ALICE
% Generation Date is 2014/Mar/11

% How to use:
%%%%%%%%% appendix with author list
%\appendix
%\section{The ALICE Collaboration}
%\label{app:collab}
%\input{authors-list.tex}  %%%%%%% get the latest version before submitting

\begingroup
\small
\begin{flushleft}
B.~Abelev\Irefn{org69}\And
J.~Adam\Irefn{org37}\And
D.~Adamov\'{a}\Irefn{org77}\And
M.M.~Aggarwal\Irefn{org81}\And
M.~Agnello\Irefn{org105}\textsuperscript{,}\Irefn{org88}\And
A.~Agostinelli\Irefn{org26}\And
N.~Agrawal\Irefn{org44}\And
Z.~Ahammed\Irefn{org124}\And
N.~Ahmad\Irefn{org18}\And
I.~Ahmed\Irefn{org15}
% ...
\renewcommand\labelenumi{\textsuperscript{\theenumi}~}

\section*{Affiliation notes}
\renewcommand\theenumi{\roman{enumi}}
\begin{Authlist}
\item \Adef{0}Deceased
\item \Adef{idp1097184}{Also at: St. Petersburg State Polytechnical University}
\item \Adef{idp2971920}{Also at: Department of Applied Physics, Aligarh Muslim University, Aligarh, India}
\item \Adef{idp3655616}{Also at: M.V. Lomonosov Moscow State University, D.V. Skobeltsyn Institute of Nuclear Physics, Moscow, Russia}
\item \Adef{idp3899264}{Also at: University of Belgrade, Faculty of Physics and "Vin\v{c}a" Institute of Nuclear Sciences, Belgrade, Serbia}
\item \Adef{idp4204816}{Permanent Address: Permanent Address: Konkuk University, Seoul, Korea}
\item \Adef{idp4745152}{Also at: Institute of Theoretical Physics, University of Wroclaw, Wroclaw, Poland}
\item \Adef{idp5650208}{Also at: University of Kansas, Lawrence, KS, United States}
\end{Authlist}

\section*{Collaboration Institutes}
\renewcommand\theenumi{\arabic{enumi}~}
\begin{Authlist}

\item \Idef{org1}A.I. Alikhanyan National Science Laboratory (Yerevan Physics Institute) Foundation, Yerevan, Armenia
\item \Idef{org2}Benem\'{e}rita Universidad Aut\'{o}noma de Puebla, Puebla, Mexico
\item \Idef{org3}Bogolyubov Institute for Theoretical Physics, Kiev, Ukraine
\item \Idef{org4}Bose Institute, Department of Physics and Centre for Astroparticle Physics and Space Science (CAPSS), Kolkata, India
\item \Idef{org5}Budker Institute for Nuclear Physics, Novosibirsk, Russia
\item \Idef{org6}California Polytechnic State University, San Luis Obispo, CA, United States
\item \Idef{org7}Central China Normal University, Wuhan, China
\item \Idef{org8}Centre de Calcul de l'IN2P3, Villeurbanne, France
\item \Idef{org9}Centro de Aplicaciones Tecnol\'{o}gicas y Desarrollo Nuclear (CEADEN), Havana, Cuba
\item \Idef{org10}Centro de Investigaciones Energ\'{e}ticas Medioambientales y Tecnol\'{o}gicas (CIEMAT), Madrid, Spain
\item \Idef{org11}Centro de Investigaci\'{o}n y de Estudios Avanzados (CINVESTAV), Mexico City and M\'{e}rida, Mexico
\item \Idef{org12}Centro Fermi - Museo Storico della Fisica e Centro Studi e Ricerche ``Enrico Fermi'', Rome, Italy
\item \Idef{org13}Chicago State University, Chicago, USA
\item \Idef{org14}Commissariat \`{a} l'Energie Atomique, IRFU, Saclay, France
\item \Idef{org15}COMSATS Institute of Information Technology (CIIT), Islamabad, Pakistan
\item \Idef{org16}Departamento de F\'{\i}sica de Part\'{\i}culas and IGFAE, Universidad de Santiago de Compostela, Santiago de Compostela, Spain
\item \Idef{org17}Department of Physics and Technology, University of Bergen, Bergen, Norway
\item \Idef{org18}Department of Physics, Aligarh Muslim University, Aligarh, India
\item \Idef{org19}Department of Physics, Ohio State University, Columbus, OH, United States
\item \Idef{org20}Department of Physics, Sejong University, Seoul, South Korea
\item \Idef{org21}Department of Physics, University of Oslo, Oslo, Norway
\item \Idef{org22}Dipartimento di Fisica dell'Universit\`{a} 'La Sapienza' and Sezione INFN Rome, Italy
\item \Idef{org23}Dipartimento di Fisica dell'Universit\`{a} and Sezione INFN, Cagliari, Italy
\item \Idef{org24}Dipartimento di Fisica dell'Universit\`{a} and Sezione INFN, Trieste, Italy
\item \Idef{org25}Dipartimento di Fisica dell'Universit\`{a} and Sezione INFN, Turin, Italy
\item \Idef{org26}Dipartimento di Fisica e Astronomia dell'Universit\`{a} and Sezione INFN, Bologna, Italy
\item \Idef{org27}Dipartimento di Fisica e Astronomia dell'Universit\`{a} and Sezione INFN, Catania, Italy
\item \Idef{org28}Dipartimento di Fisica e Astronomia dell'Universit\`{a} and Sezione INFN, Padova, Italy
\item \Idef{org29}Dipartimento di Fisica `E.R.~Caianiello' dell'Universit\`{a} and Gruppo Collegato INFN, Salerno, Italy
\item \Idef{org30}Dipartimento di Scienze e Innovazione Tecnologica dell'Universit\`{a} del  Piemonte Orientale and Gruppo Collegato INFN, Alessandria, Italy
\item \Idef{org31}Dipartimento Interateneo di Fisica `M.~Merlin' and Sezione INFN, Bari, Italy
\item \Idef{org32}Division of Experimental High Energy Physics, University of Lund, Lund, Sweden
\item \Idef{org33}Eberhard Karls Universit\"{a}t T\"{u}bingen, T\"{u}bingen, Germany
\item \Idef{org34}European Organization for Nuclear Research (CERN), Geneva, Switzerland
\item \Idef{org35}Faculty of Engineering, Bergen University College, Bergen, Norway
\item \Idef{org36}Faculty of Mathematics, Physics and Informatics, Comenius University, Bratislava, Slovakia
\item \Idef{org37}Faculty of Nuclear Sciences and Physical Engineering, Czech Technical University in Prague, Prague, Czech Republic
\item \Idef{org38}Faculty of Science, P.J.~\v{S}af\'{a}rik University, Ko\v{s}ice, Slovakia
\item \Idef{org39}Frankfurt Institute for Advanced Studies, Johann Wolfgang Goethe-Universit\"{a}t Frankfurt, Frankfurt, Germany
\item \Idef{org40}Gangneung-Wonju National University, Gangneung, South Korea
\item \Idef{org41}Gauhati University, Department of Physics, Guwahati, India
\item \Idef{org42}Helsinki Institute of Physics (HIP), Helsinki, Finland
\item \Idef{org43}Hiroshima University, Hiroshima, Japan
\item \Idef{org44}Indian Institute of Technology Bombay (IIT), Mumbai, India
\item \Idef{org45}Indian Institute of Technology Indore, Indore (IITI), India
\item \Idef{org46}Institut de Physique Nucl\'eaire d'Orsay (IPNO), Universit\'e Paris-Sud, CNRS-IN2P3, Orsay, France
\item \Idef{org47}Institut f\"{u}r Informatik, Johann Wolfgang Goethe-Universit\"{a}t Frankfurt, Frankfurt, Germany
\item \Idef{org48}Institut f\"{u}r Kernphysik, Johann Wolfgang Goethe-Universit\"{a}t Frankfurt, Frankfurt, Germany
\item \Idef{org49}Institut f\"{u}r Kernphysik, Westf\"{a}lische Wilhelms-Universit\"{a}t M\"{u}nster, M\"{u}nster, Germany
\item \Idef{org50}Institut Pluridisciplinaire Hubert Curien (IPHC), Universit\'{e} de Strasbourg, CNRS-IN2P3, Strasbourg, France
\item \Idef{org51}Institute for Nuclear Research, Academy of Sciences, Moscow, Russia
\item \Idef{org52}Institute for Subatomic Physics of Utrecht University, Utrecht, Netherlands
\item \Idef{org53}Institute for Theoretical and Experimental Physics, Moscow, Russia
\item \Idef{org54}Institute of Experimental Physics, Slovak Academy of Sciences, Ko\v{s}ice, Slovakia
\item \Idef{org55}Institute of Physics, Academy of Sciences of the Czech Republic, Prague, Czech Republic
\item \Idef{org56}Institute of Physics, Bhubaneswar, India
\item \Idef{org57}Institute of Space Science (ISS), Bucharest, Romania
\item \Idef{org58}Instituto de Ciencias Nucleares, Universidad Nacional Aut\'{o}noma de M\'{e}xico, Mexico City, Mexico
\item \Idef{org59}Instituto de F\'{\i}sica, Universidad Nacional Aut\'{o}noma de M\'{e}xico, Mexico City, Mexico
\item \Idef{org60}iThemba LABS, National Research Foundation, Somerset West, South Africa
\item \Idef{org61}Joint Institute for Nuclear Research (JINR), Dubna, Russia
\item \Idef{org62}Korea Institute of Science and Technology Information, Daejeon, South Korea
\item \Idef{org63}KTO Karatay University, Konya, Turkey
\item \Idef{org64}Laboratoire de Physique Corpusculaire (LPC), Clermont Universit\'{e}, Universit\'{e} Blaise Pascal, CNRS--IN2P3, Clermont-Ferrand, France
\item \Idef{org65}Laboratoire de Physique Subatomique et de Cosmologie, Universit\'{e} Grenoble-Alpes, CNRS-IN2P3, Grenoble, France
\item \Idef{org66}Laboratori Nazionali di Frascati, INFN, Frascati, Italy
\item \Idef{org67}Laboratori Nazionali di Legnaro, INFN, Legnaro, Italy
\item \Idef{org68}Lawrence Berkeley National Laboratory, Berkeley, CA, United States
\item \Idef{org69}Lawrence Livermore National Laboratory, Livermore, CA, United States
\item \Idef{org70}Moscow Engineering Physics Institute, Moscow, Russia
\item \Idef{org71}National Centre for Nuclear Studies, Warsaw, Poland
\item \Idef{org72}National Institute for Physics and Nuclear Engineering, Bucharest, Romania
\item \Idef{org73}National Institute of Science Education and Research, Bhubaneswar, India
\item \Idef{org74}Niels Bohr Institute, University of Copenhagen, Copenhagen, Denmark
\item \Idef{org75}Nikhef, National Institute for Subatomic Physics, Amsterdam, Netherlands
\item \Idef{org76}Nuclear Physics Group, STFC Daresbury Laboratory, Daresbury, United Kingdom
\item \Idef{org77}Nuclear Physics Institute, Academy of Sciences of the Czech Republic, \v{R}e\v{z} u Prahy, Czech Republic
\item \Idef{org78}Oak Ridge National Laboratory, Oak Ridge, TN, United States
\item \Idef{org79}Petersburg Nuclear Physics Institute, Gatchina, Russia
\item \Idef{org80}Physics Department, Creighton University, Omaha, NE, United States
\item \Idef{org81}Physics Department, Panjab University, Chandigarh, India
\item \Idef{org82}Physics Department, University of Athens, Athens, Greece
\item \Idef{org83}Physics Department, University of Cape Town, Cape Town, South Africa
\item \Idef{org84}Physics Department, University of Jammu, Jammu, India
\item \Idef{org85}Physics Department, University of Rajasthan, Jaipur, India
\item \Idef{org86}Physik Department, Technische Universit\"{a}t M\"{u}nchen, Munich, Germany
\item \Idef{org87}Physikalisches Institut, Ruprecht-Karls-Universit\"{a}t Heidelberg, Heidelberg, Germany
\item \Idef{org88}Politecnico di Torino, Turin, Italy
\item \Idef{org89}Purdue University, West Lafayette, IN, United States
\item \Idef{org90}Pusan National University, Pusan, South Korea
\item \Idef{org91}Research Division and ExtreMe Matter Institute EMMI, GSI Helmholtzzentrum f\"ur Schwerionenforschung, Darmstadt, Germany
\item \Idef{org92}Rudjer Bo\v{s}kovi\'{c} Institute, Zagreb, Croatia
\item \Idef{org93}Russian Federal Nuclear Center (VNIIEF), Sarov, Russia
\item \Idef{org94}Russian Research Centre Kurchatov Institute, Moscow, Russia
\item \Idef{org95}Saha Institute of Nuclear Physics, Kolkata, India
\item \Idef{org96}School of Physics and Astronomy, University of Birmingham, Birmingham, United Kingdom
\item \Idef{org97}Secci\'{o}n F\'{\i}sica, Departamento de Ciencias, Pontificia Universidad Cat\'{o}lica del Per\'{u}, Lima, Peru
\item \Idef{org98}Sezione INFN, Bari, Italy
\item \Idef{org99}Sezione INFN, Bologna, Italy
\item \Idef{org100}Sezione INFN, Cagliari, Italy
\item \Idef{org101}Sezione INFN, Catania, Italy
\item \Idef{org102}Sezione INFN, Padova, Italy
\item \Idef{org103}Sezione INFN, Rome, Italy
\item \Idef{org104}Sezione INFN, Trieste, Italy
\item \Idef{org105}Sezione INFN, Turin, Italy
\item \Idef{org106}SSC IHEP of NRC Kurchatov institute, Protvino, Russia
\item \Idef{org107}SUBATECH, Ecole des Mines de Nantes, Universit\'{e} de Nantes, CNRS-IN2P3, Nantes, France
\item \Idef{org108}Suranaree University of Technology, Nakhon Ratchasima, Thailand
\item \Idef{org109}Technical University of Split FESB, Split, Croatia
\item \Idef{org110}The Henryk Niewodniczanski Institute of Nuclear Physics, Polish Academy of Sciences, Cracow, Poland
\item \Idef{org111}The University of Texas at Austin, Physics Department, Austin, TX, USA
\item \Idef{org112}Universidad Aut\'{o}noma de Sinaloa, Culiac\'{a}n, Mexico
\item \Idef{org113}Universidade de S\~{a}o Paulo (USP), S\~{a}o Paulo, Brazil
\item \Idef{org114}Universidade Estadual de Campinas (UNICAMP), Campinas, Brazil
\item \Idef{org115}University of Houston, Houston, TX, United States
\item \Idef{org116}University of Jyv\"{a}skyl\"{a}, Jyv\"{a}skyl\"{a}, Finland
\item \Idef{org117}University of Liverpool, Liverpool, United Kingdom
\item \Idef{org118}University of Tennessee, Knoxville, TN, United States
\item \Idef{org119}University of Tokyo, Tokyo, Japan
\item \Idef{org120}University of Tsukuba, Tsukuba, Japan
\item \Idef{org121}University of Zagreb, Zagreb, Croatia
\item \Idef{org122}Universit\'{e} de Lyon, Universit\'{e} Lyon 1, CNRS/IN2P3, IPN-Lyon, Villeurbanne, France
\item \Idef{org123}V.~Fock Institute for Physics, St. Petersburg State University, St. Petersburg, Russia
\item \Idef{org124}Variable Energy Cyclotron Centre, Kolkata, India
\item \Idef{org125}Vestfold University College, Tonsberg, Norway
\item \Idef{org126}Warsaw University of Technology, Warsaw, Poland
\item \Idef{org127}Wayne State University, Detroit, MI, United States
\item \Idef{org128}Wigner Research Centre for Physics, Hungarian Academy of Sciences, Budapest, Hungary
\item \Idef{org129}Yale University, New Haven, CT, United States
\item \Idef{org130}Yonsei University, Seoul, South Korea
\item \Idef{org131}Zentrum f\"{u}r Technologietransfer und Telekommunikation (ZTT), Fachhochschule Worms, Worms, Germany
\end{Authlist}
\endgroup